\documentclass[10pt,journal,compsoc]{IEEEtran}

\usepackage{caption}
\usepackage{cite}
\usepackage{optidef}
\usepackage{graphicx}
\usepackage{algorithm2e}
\usepackage{algpseudocode}
\usepackage{amsmath}
\usepackage{svg}
\usepackage{tabularx}

\newif\ifblackandwhite

\usepackage{booktabs}
\usepackage{subcaption}
\usepackage{hyperref}
\definecolor{lightgray}{gray}{0.9}
\usepackage{color, colortbl}
\definecolor{Gray}{gray}{0.9}
             
\begin{document}

\title{What are the real implications for $\text{CO}_2$ as generation from renewables increases?}

\author{Dhruv~Suri,
        Jacques~de~Chalendar,
        and~Ines~M.~L.~Azevedo
}

\IEEEtitleabstractindextext{%
\begin{abstract}
Wind and solar electricity generation account for 14\% of total electricity generation in the United States and are expected to continue to grow in the next decades. In low carbon systems, generation from renewable energy sources displaces conventional fossil fuel power plants resulting in lower system-level emissions and emissions intensity. However, we find that intermittent generation from renewables changes the way conventional thermal power plants operate, and that the displacement of generation is not 1:1 as expected; as renewable generation increases, the rest of the system needs to cope with variability, leading to suboptimal operation levels for thermal power plants. We use hourly data at the power plant level for the California Independent System Operator and for the Electricity Reliability Council of Texas to estimate the magnitude of the emissions' penalty associated with renewable variability. We find that the emissions intensity of conventional generation is up to 10x higher at the lower capacity factors that several power plants now operate at. However, this results in a very small effect in terms of inducing emissions. Indeed, solar generation in California displaces roughly 91\% of the emissions that would have been displaced without the increased emissions intensity in the state’s thermal power plants. In Texas, the analogous figure for wind generation is 95\%. Despite this effect being small under current level of renewable generation, policy maker ought to continue to monitor potential emissions penalties as variable renewable energy sources continue to increase. Our work provide a method that allows policy and decision makers to continue to track the effect of renewable capacity additional and the resulting thermal power plant operational responses.

\end{abstract}

}

\maketitle

\IEEEdisplaynontitleabstractindextext

\section{Introduction}
\label{sec:introduction}

Deep decarbonization of the electric power sector is needed in order to address climate change and to move energy systems towards sustainability. The decarbonization of electricity in most regions requires an accelerated and large-scale deployment of renewable energy, such as solar and wind \cite{portner2022climate}. Many regions have ambitious goals for decarbonization. In this study, we focus specifically on the California Independent System Operator (CAISO) and on the Electric Reliability Council of Texas (ERCOT). This focus is motivated by the fact that they have a large share of renewable energy sources (RES) and represent the largest annual state electricity demand. California and Texas, which are the focus of this analysis, have specific legislation on decarbonization targets \cite{SB100,choukulkar2022don,yuan2020challenges, di2019synthesis, zhu2022decarbonization}. Senate Bill (SB) 100 established a landmark policy in California  requiring renewable energy and zero-carbon resources supply 100\% of electric retail sales to end-use customers by 2045\cite{baik2022california}. Texas was one of the first states to adopt a renewable portfolio standard (RPS) and uses a market-based approach to scale renewable energy in the state \cite{bowman2020renewable}.

By 2021, Texas and California accounted for the largest and second largest states in terms of electricity demand \cite{EIA_CA}. Total electricity consumption in California has been declining \cite{CEC_CA}, in part owing to energy efficiency and conservation \cite{EIA_CA}. Simultaneously, wind and solar-PV capacity (and generation) has been increasing in both states, now accounting for 28.4\% of annual generation in California and 25.4\% in Texas (see Table \ref{tab:summary})\cite{EIA_monthly}. In 2010, California's generation mix consisted of 7 coal plants \cite{CA_coal} and 213 gas plants \cite{EGRID_2010}, with only 2.9\% and 0.4\% of net generation from wind and solar. By 2021, 7.7\% and 17.7\% of the state’s net generation was from wind and solar, respectively \cite{EGRID}. 
 
Similarly, Texas also consistently ramped up its share of electricity production from renewables. In 2010, ERCOT's share of generation from wind and solar accounted for 6.35\% and 0.002\% of the balancing authority's (BA) net electricity generation. In 2021, the share of generation from wind and solar increased to 20.7\% and 3.1\% respectively.

\begin{table*}[ht]
    \centering
    \renewcommand{\arraystretch}{1.2} 
    \caption{Regional sources of electricity generation, emissions, and emissions intensity for CAISO and ERCOT. Data obtained from the United States Environmental Protection Agency's (EPA) Emissions and Generation Resource Integrated Database (eGRID). \cite{epa2024emissions}}
    \label{tab:summary}
    \begin{tabular}{>{\raggedright\arraybackslash}p{0.42\textwidth} *{6}{>{\centering\arraybackslash}p{0.05\textwidth}}}
        \toprule
        & \multicolumn{3}{c}{\textbf{California (CAISO)}} & \multicolumn{3}{c}{\textbf{Texas (ERCOT)}} \\
        \cmidrule(lr){2-4} \cmidrule(lr){5-7}
        & \textbf{2010} & \textbf{2015} & \textbf{2021} & \textbf{2010} & \textbf{2015} & \textbf{2021} \\
        \midrule
        \rowcolor{lightgray}
        \textbf{Natural gas} & & & & & & \\
        Number of plants & 213 & 286 & 298 & 116 & 165 & 216 \\
        Installed capacity (GW) & 38.28 & 49.61 & 33.83 & 63.37 & 89.97 & 63.41 \\
        Share of generation (\%) & 51.32 & 47.94 & 47.39 & 45.23 & 46.97 & 45.28 \\
        Share of emissions (\%) & 88.79 & 94.98 & 95.42 & 32.83 & 39.41 & 48.28 \\
        Average emissions intensity (t$\text{CO}_2$/MWh) & 0.5 & 0.47 & 0.55 & 0.49 & 0.47 & 0.58 \\
        Std dev of emissions intensity (t$\text{CO}_2$/MWh) & 0.63 & 0.33 & 1.27 & 0.18 & 0.20 & 0.20 \\
        Average capacity factor (\%) & 0.39 & 0.33 & 0.41 & 0.32 & 0.31 & 0.20 \\
        Std dev of capacity factor (\%) & 0.33 & 0.31 & 0.34 & 0.25 & 0.25 & 0.24 \\
        \midrule
        \rowcolor{lightgray}
        \textbf{Coal} & & & & & & \\
        Number of plants & 7 & 3 & 1 & 16 & 16 & 11 \\
        Installed capacity (GW) & 2.01 & 1.74 & 0.06 & 21.09 & 26.33 & 16.53 \\
        Share of generation (\%) & 1.32 & 0.19 & 0.18 & 35.35 & 26.03 & 18.37 \\
        Share of emissions (\%) & 4.82 & 0.52 & 0.45 & 66.74 & 59.49 & 50.82 \\
        Average emissions intensity (t$\text{CO}_2$/MWh) & 0.69 & 0.53 & 0.52 & 1.08 & 1.06 & 0.96 \\
        Std dev of emissions intensity (t$\text{CO}_2$/MWh) & 0.38 & - & - & 0.27 & 0.09 & 0.33 \\
        Average capacity factor (\%) & 0.62 & 0.6 & 0.56 & 0.69 & 0.47 & 0.54 \\
        Std dev of capacity factor (\%) & 0.29 & - & - & 0.18 & 0.14 & 0.22 \\
        \midrule
        \rowcolor{lightgray}
        \textbf{Utility scale solar} & & & & & & \\
        Number of plants & 0 & 483 & 595 & 0 & 39 & 104 \\
        Installed capacity (GW) & 0.0 & 14.37 & 15.53 & 0.0 & 2.82 & 8.92 \\
        Share of generation (\%) & 0.0 & 11.22 & 19.5 & 0.0 & 0.18 & 3.53 \\
        Average capacity factor (\%) & - & 0.22 & 0.21 & - & 0.17 & 0.18 \\
        Std dev of capacity factor (\%) & - & 0.09 & 0.09 & - & 0.08 & 0.08 \\
        \midrule
        \rowcolor{lightgray}
        \textbf{Utility scale wind} & & & & & & \\
        Number of plants & 0 & 129 & 108 & 0 & 146 & 169 \\
        Installed capacity (GW) & 0.0 & 6.17 & 6.15 & 0.0 & 26.39 & 32.23 \\
        Share of generation (\%) & 0.0 & 7.92 & 8.91 & 0.0 & 13.75 & 21.90 \\
        Average capacity factor (\%) & - & 0.25 & 0.27 & - & 0.32 & 0.32 \\
        Std dev of capacity factor (\%) & - & 0.1 & 0.11 & - & 0.10 & 0.11 \\
        \bottomrule
    \end{tabular}
\end{table*}

In this study, we posit that an  effect of increased generation from wind and solar is the lowering of overall emissions and the emissions intensity of the electricity system but this displacement is not 1:1 as expected: As renewable generation increases, the rest of the system needs to cope with variability. The magnitude of the effects associated with this variability are not well understood \cite{fell2021regional, katzenstein2009air}. We focus on estimating these effects using data from historical power system operation for CAISO and for ERCOT over the last ten years.

Previously, authors have used simulation and regression based approaches to understand the effect of RES on thermal generators, making important contributions that inform our modelling framework. Katzenstein and Apt simulate a system of wind, solar, and gas and assess the emissions from two types of natural gas generators \cite{katzenstein2009air}. They find that steam injection gas generators achieve only 30-50\% of expected $\text{NO}_x$ emissions reductions and close to 80\% of expected $\text{CO}_2$ emissions reductions. Given the limited data availability for turbine level data at the time, the authors limit their scope to  two types of natural gas generators. 

Others, like Eser et al. \cite{eser2016effect}, show the effect of increased penetration of RES on thermal power plants in Central Western and Eastern Europe. Through a power flow model, the authors show that the penetration of RES  induces a 4-23\% increase in the number of starts in conventional plants, and a 63-181\% increase in load ramp. Eser et al. \cite{eser2016effect} findings are similar to those from other studies regarding the effect of cycling on power plant operational life: not only does increased cycling result in more wear and tear on gas generators \cite{lefton1995managing,kumar2012power,lew2012impacts} but also reduces emissions performance, especially at lower capacity factors \cite{macak2005evaluation}. Valentino et al. show the change in the number of startups and the degradation of plant heat rate with lower utilization \cite{valentino2012system}. The authors use a dispatch model to determine active units, and estimate a fuel consumption function based on different blocks of heat rate data. However, the study does not consider heat rate as a function of plant type, ramping, vintage, and degree of cycling, which is a key determinant of heat rates at low utilization levels. Other relevant studies include Makarov \cite{makarov2009operational}, who studied the operational impact of wind generation in CAISO. They concluded that wind generation has limited effects on load following and regulation within CAISO's operational region. 

Some studies focus on how RES may change the unit commitment and economic dispatch of generators \cite{vithayasrichareon2017operational,goransson2017impact,hlalele2020dynamic} and conclude that increasing wind and solar changes market dynamics in both nodal and wholesale electricity market designs. Other studies find that RES lead to thermal generators increase in ramping, lower utilization, and higher operations and maintenance costs \cite{cui2017characterizing, shahmohammadi2018role, lamadrid2012ancillary, van2013impact, de2020estimating}.

While unit commitment and economic dispatch studies using simulations provide valuable insights and demonstrate that increasing penetration of RES changes several operational characteristics of thermal plants, these approaches may miss real world operations in the electricity system, including coping with transmission constraints, with policies, and with RES variability  \cite{fell2021regional}. Empirical approaches that rely on observational  data mitigate some of these limitations, but may not capture future conditions with larger RES capacity and generation. Examples of such approaches include the studies by \cite{fell2021regional}, \cite{mills2020impacts}, \cite{wiser2017impacts}. These studies find that renewables do indeed displace thermal generation, however, the exact nature of the displacement is still heterogenous across the studies.

\begin{figure*}[ht!]
    \centering
    \includegraphics[width=1\linewidth]{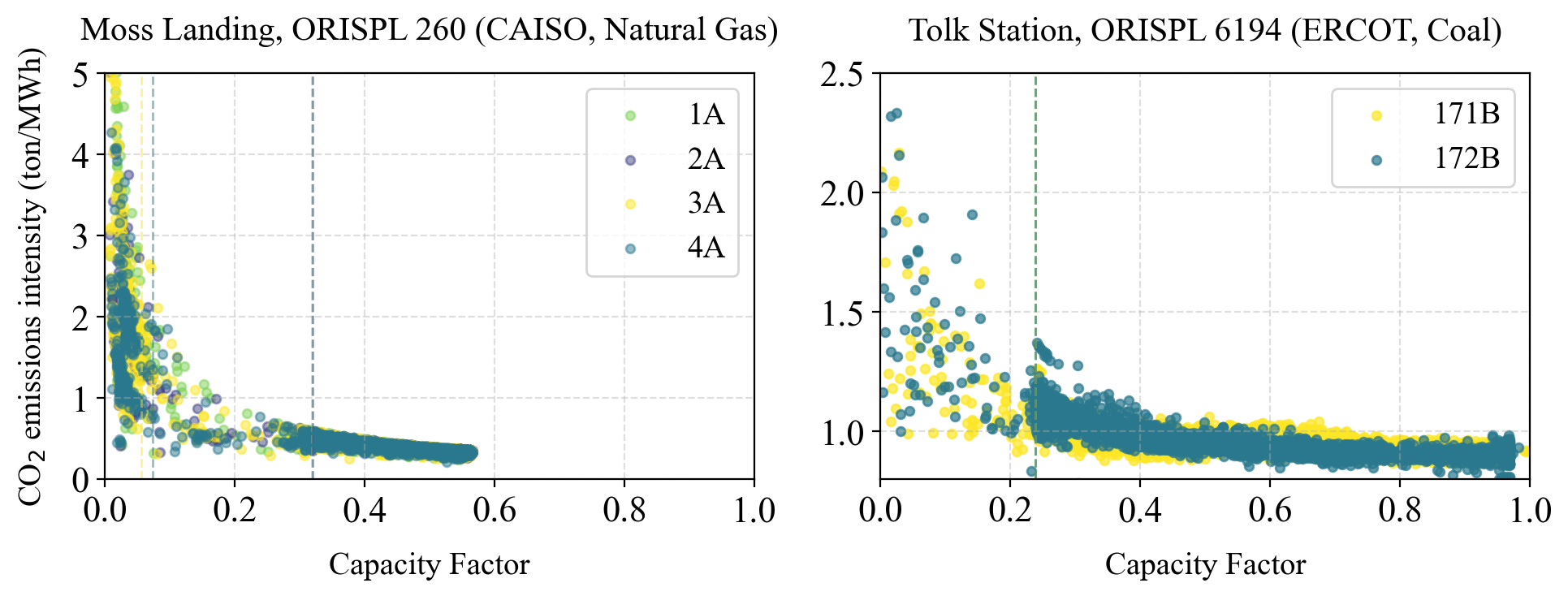}
    \caption{Emissions intensity versus capacity factor for two sample gas plants. Moss Landing is a natural gas-fired plant in CAISO, whereas Tolk Station is a coal power plant in ERCOT. Colors differentiate individual units, and the dashed lines indicate P10 for the capacity factor for each unit}
    \label{fig:sample_CF}
\end{figure*}

To date, no study as far as we are aware of has looked longitudinally and empirically at the implications for emissions and changes in ramping for thermal power plants associated with renewables in the electricity grid, and how the increased emissions intensity can be abated given more efficient dispatch constraints. We provide a framework that allows to assess the implications of renewables on thermal plant operations, and present an analysis of historical thermal generation data from 2018 through 2023 for CAISO and ERCOT. The rest of this paper is organized as follows. First we compare what the emissions are under different simple operational assumptions. We then we use historical data to estimate via regression the system-level change in generation and emissions with a 1\% increase in generation from wind and solar, and dive deeper into individual plant-level response. Lastly, we characterize the effect of increased intermittency as generation from renewables electricity generation increases.

\section{Results}

\subsection{Comparing three emissions intensity scenarios}

\begin{figure*}[ht!]
    \centering
    \includegraphics[width=1\linewidth]{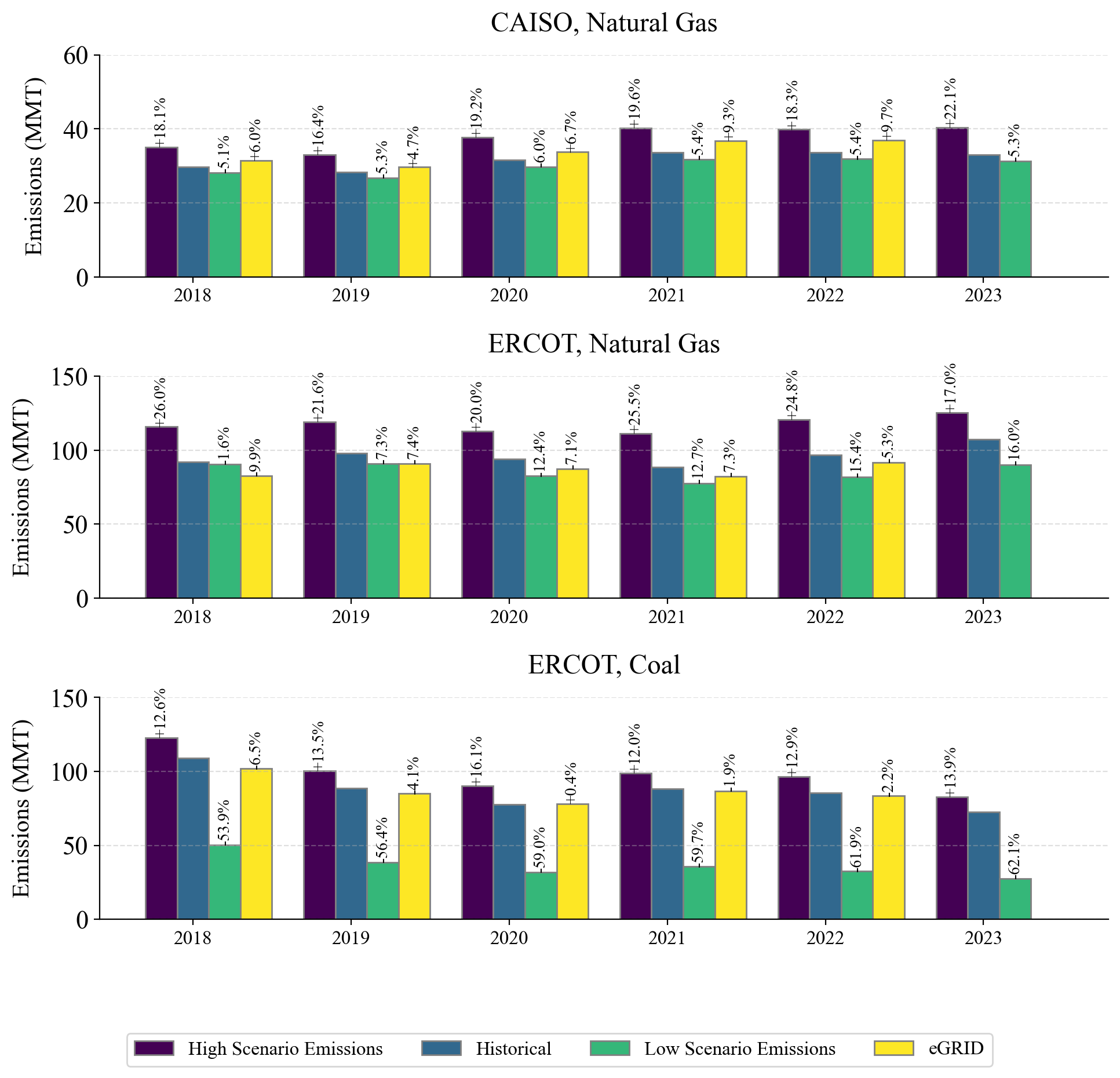}
    \caption{Three emissions intensity scenarios that represent different operational regimes of thermal plants: (i) a ‘high emissions’ scenario where we consider plants operating at the 90th percentile of their emissions intensity, (ii) a ‘low emissions’ scenario where we consider the observed emissions intensity of the plant at the 10th percentile of their observed emissions intensity, and (iii) an ‘eGRID’ scenario where we consider plants consistently operating at an emissions intensity specified for the plant category in eGRID for the respective reporting year.}
    \label{fig:energy_sources_comparison}
\end{figure*}

We use hourly thermal generation and $\text{CO}_2$ emissions from CEMS at the generating unit and power plant level to study the operational characteristics of power plants. We start by assessing the observed emissions in each year from 2018 to 2023 in CAISO and ERCOT, and then compare those emissions to what they would have been under different operational scenarios. Specifically, we contrast the emissions observed historically to those that would arise if the power plants were operating under an emissions intensity at a low capacity factor, as well as that at a high capacity factor observed for each plant. We also compare such emissions to those reported by each plant in eGRID. 

The motivation for these simple scenarios is explained by Figure \ref{fig:sample_CF}. Using CEMS data, we have observed that the emissions intensity of power plants changes  markedly as the capacity factor changes. At lower capacity factors, plants consume more fuel per unit of electricity generated - and thus have a higher emissions' intensity - because they are not operating at their ideal operational levels. Figure \ref{fig:sample_CF} illustrates this relationship between emissions intensity and capacity factor by generating unit for two plants in CAISO and ERCOT. Each data point in the figure represents an hourly operational value at which the generation and cumulative emissions were measured. In both instances, we see that the emissions intensity at lower capacity factors is several multiples of the baseline emissions intensity at higher utilization. For Moss Landing, a natural gas plant in CAISO, the mean emissions intensity for a capacity factor greater than 0.3 is 0.36 ton$\text{CO}_2$/MWh. In contrast, when the capacity factor is less that 0.05, the emissions intensity of the plant is 0.84 ton$\text{CO}_2$/MWh . Tolk Station, a coal-fired plant in ERCOT, generally serves as a baseload plant. The mean emissions intensity for a capacity factor greater than 0.3 is 0.94 ton$\text{CO}_2$/MWh whereas when the capacity factor is less than 0.05, the mean emissions intensity is 1.426 ton$\text{CO}_2$/MWh. This increase in emissions is not unique to the two plants in question, but is seen across the entire fleet of thermal plants in CAISO and ERCOT. In the SI, Section \ref{EI_CF_plots}, we show similar plots for every single natural gas and coal plant in CAISO and ERCOT. 

As the generation from renewable energy sources increases, more thermal plants will be required to operate at low capacity factors or will remain online as spinning reserves given ramping and startup constraints (see Figure \ref{fig:hourly_metrics-1} in Section \ref{change_in_op} of the SI). Low utilization will in turn result in higher hourly heat rates as a direct consequence of which system-level emissions will also increase. To model these effects, we start by considering three simple emissions intensity scenarios that represent different operational regimes of thermal plants: (i) a ‘high emissions’ scenario where we select, for each plant, it's 90th percentile of emissions intensity, (ii) a ‘low emissions’ scenario where we consider the observed emissions intensity of the plant at the 10th percentile of their observed emissions intensity in CEMS, and (iii) an ‘eGRID’ scenario where we consider plants consistently operating at an emissions intensity specified for that respective plants in eGRID for the respective reporting year. We do not specify an eGRID scenario for 2023 since data is released with a lag of one year. Of course, these high and low emissions intensity scenarios are related to how the plants are operated, and thus in the next results sections we will address this issue using regression based models. Figure \ref{fig:energy_sources_comparison} shows the resulting annual emissions with a high, low and observed emissions intensity scenario, along with a comparison with eGRID plant-level emissions. These simple scenarios provide important bounding results: if power plants were to be operating far from their optimal operation limits, i.e., with low capacity facts and high emissions intensity, the overall $\text{CO}_2$ emissions from natural gas and coal power plants in CAISO and ERCOT would be somewhere between 12\% and 26\% higher than the observed emissions (depending on the year).  

\subsection{Marginal impacts of wind and solar on plant-level generation, emissions and emissions intensity}

\begin{table*}[ht!]
\centering
\caption{Coefficients for the panel regression formulation performed on hourly data aggregated to daily time series for CAISO (left) and ERCOT (right) where each of the columns represent the coefficients when generation, emissions, and emissions intensity are the endogenous variable, respectively.}
\label{table:coefficients}

\begin{minipage}{0.5\textwidth}
    \centering
    \begin{tabular}{cccc}
        \toprule
        \rowcolor{Gray} Coefficient & Generation & Emissions & Intensity \\
        \midrule
        Thermal generation & 2.80*** & 2.66*** & -0.14*** \\
        & (0.068) & (0.062) & (0.010) \\
        Solar & -0.22*** & -0.21*** & 0.02*** \\
        & (0.031) & (0.028) & (0.004) \\
        Wind & -0.20*** & -0.19*** & 0.01***\\
        & (0.012) & (0.011) & (0.002) \\
        Wind ramp & 0.12*** & 0.11*** & -0.004\\
        & (0.015) & (0.014) & (0.002)\\
        Solar (ext) & 0.004 & 0.003 & -0.001 \\
        & (0.031) & (0.029) & (0.004)\\
        Wind (ext) & -0.02*** & -0.02* & 0.002*** \\
        & (0.005) & (0.004) & (0.001)\\
        \midrule
        R-squared & 0.84 & 0.84 & 0.77 \\
        No. of obs & 44,724 & 44,724 & 44,724 \\
        \bottomrule
    \end{tabular}
    \label{tab:reg-california}
\end{minipage}\hfill
\begin{minipage}{0.5\textwidth}
    \centering
    \begin{tabular}{cccc}
        \toprule
        \rowcolor{Gray} Coefficient & Generation & Emissions & Intensity \\
        \midrule
        Thermal generation & 1.78*** & 1.67*** & -0.11*** \\
        & (0.099) & (0.093) & (0.014) \\
        Solar & -0.03 & -0.03* & 0.003 \\
        & (0.016) & (0.015) & (0.002) \\
        Wind & -0.33*** & -0.31*** & 0.03***\\
        & (0.019) & (0.018) & (0.003) \\
        Wind ramp & 0.07*** & 0.07*** & 0.001\\
        & (0.020) & (0.019) & (0.003)\\
        Solar (ext) & 0.05*** & 0.06*** & -0.001 \\
        & (0.015) & (0.014) & (0.002)\\
        Wind (ext) & -0.03** & -0.03 & -0.002 \\
        & (0.019) & (0.018) & (0.003)\\
        \midrule
        R-squared & 0.77 & 0.80 & 0.92 \\
        No. of obs & 87,089 & 87,089 & 87,089 \\
        \bottomrule
    \end{tabular}
    \label{tab:reg-texas}
\end{minipage}
\end{table*}

Now, we turn to provide more accurate estimates the effect that from renewables on thermal power plants' operations using the regression model described in the data and methods section. To attribute the impact of renewables on thermal plants, we use a fixed-effects panel regression model with a logarithmic specification. Instead of considering generation from renewables as a whole, we separate solar and wind generation as discrete independent variables given their varying levels of penetration in CAISO and ERCOT. The logarithmic specification used is given by Equation \ref{log_spec_1}.The coefficients of this model can thus be interpreted as the percent change in the dependent variable (emissions intensity, total emissions and generation), per unit percent increase in generation from wind and solar. The model accounts for location and time fixed effects which allows us to interpret the coefficients independently from the plant fuel type.

In Table \ref{table:coefficients}, we show the coefficients of the panel regression model for the two regions. In this way, we are able to quantify the percent change in emissions intensity, total emissions, and generation for thermal power plants attributable to a marginal increase in generation from solar and wind.  Here, the term ‘marginal’ implies an additional unit (or unit percent) increase in generation from renewables and its resulting impact. This in turn can be used to determine the expected vs actual emissions displacement from renewables (see Section \ref{deviation}).

As shown in Table \ref{table:coefficients}, an increase in wind and solar consistently results in lower thermal generation and lower emissions. That said, if this displacement was 100\% efficient (or 1:1), both generation and emissions would reduce by an equivalent amount, and thus should have identical coefficients in the fixed-effects model. In CAISO, the displacement from wind and solar is unequal. A 1\% increase in generation from solar results in a 0.22\% reduction in thermal generation and 0.20\% decrease in total emissions. Thus, increased generation from solar achieves 91.3\% of its expected emissions displacement (see the methods section). Similarly, wind achieves 95.0\% of its expected emissions displacement. In ERCOT, on the other hand, displacement from wind is an order of magnitude more pronounced than the displacement from solar, which may be attributable to the difference in the installed utility-scale capacity of the two resources in the BA. Looking at the displacement induced from wind, we find that a 1\% increase in generation results in a 0.33\% reduction in thermal generation and 0.31\% decrease in total emissions. Thus, wind generation in ERCOT achieves 91.2\% of its expected emissions displacement. 

While the expected vs actual emissions displacement helps us understand the system-level impact of renewables, it still does not enable us to attribute the effect of renewables on plant-level emissions intensity. Looking closely at the coefficients of the model with emissions intensity as the dependent variable, we find that a 1\% increased generation from wind and solar does indeed impact plant-level emissions intensity for existing thermal generators. For CAISO, thermal power plant intensities are more susceptible to increased generation from solar as compared to wind (0.02\% for solar, 0.01\% for wind). In ERCOT, generation from wind results in a 0.03\% increase in emissions intensity of plants. While at first, these numbers may seem insignificant, the hourly ramp in solar generation in CAISO during the day can reach ~50\%, which would result in a 1\% increase in the emissions intensity of power plants for that hour. If we consider CAISO’s 2022 system-level emissions averaged uniformly across all hours, a 1\% increase in intensity across all power plants  is equivalent to 53 tons of $\text{CO}_2$. As the ramp becomes more pronounced with increasing intermittent renewable capacity, this marginal increase should not be ignored.

Intermittency in generation from wind is measured by the daily ‘wind ramp’ variable, represented by $\overline{W}_t$ in the model specification. In contrast to the coefficients for wind and solar generation, an increase in the variability of the wind results in a subsequent increase in generation from thermal plants in both regions. The magnitude of the positive displacement is also quite significant - 52\% for CAISO, and 35\% for ERCOT. Clearly, wind and solar generation play a key role in determining the generation and emissions in thermal plants, however, variability and intermittency in the generation from renewables has a significant impact as well.

\begin{figure*}
    \centering
    \includegraphics[width=1\linewidth]{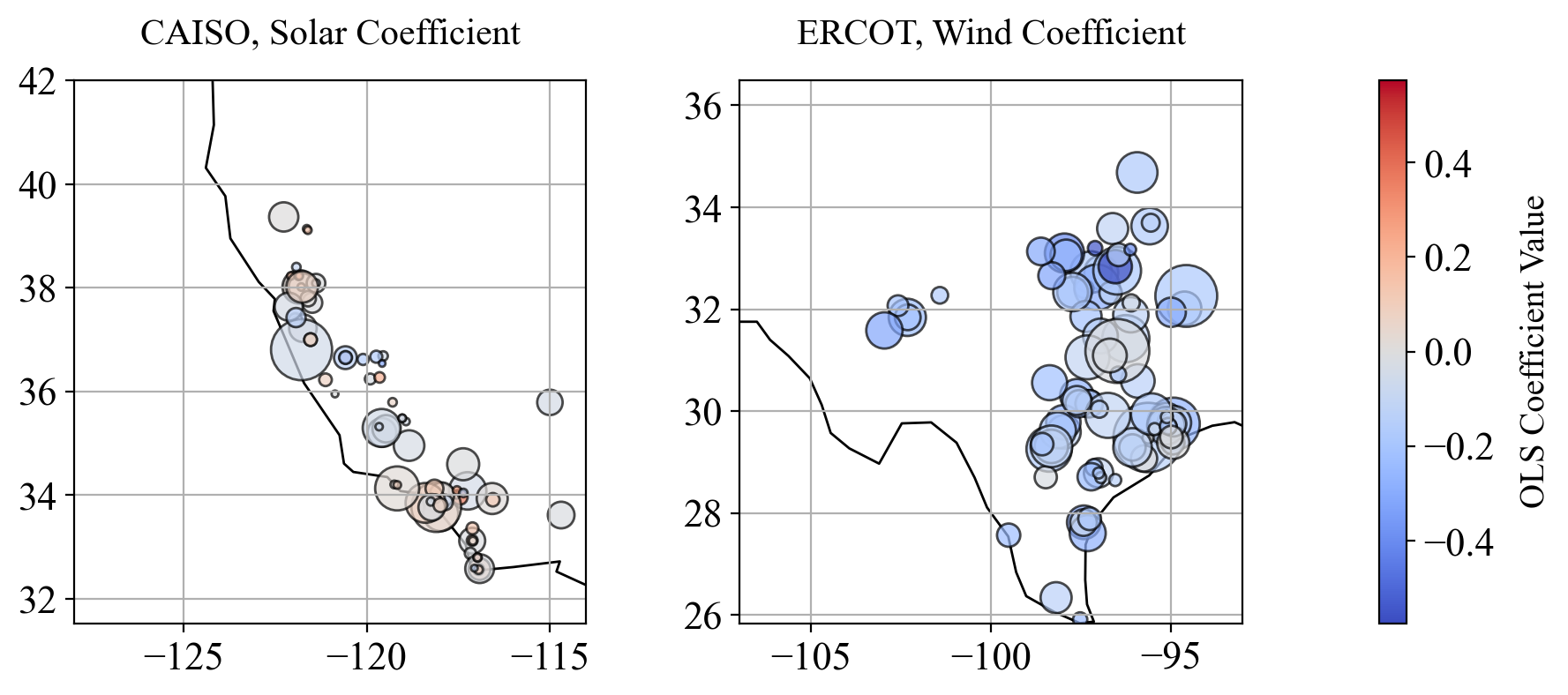}
    \caption{Individual plant regression OLS estimates for generation from natural gas plants in CAISO in response to solar (left) and coal and natural gas plants in ERCOT in response to wind (right). The size of the points indicates the nameplate capacity.}
    \label{fig:map-coeff}
\end{figure*}

\begin{figure*}
    \centering
    \includegraphics[width=1\linewidth]{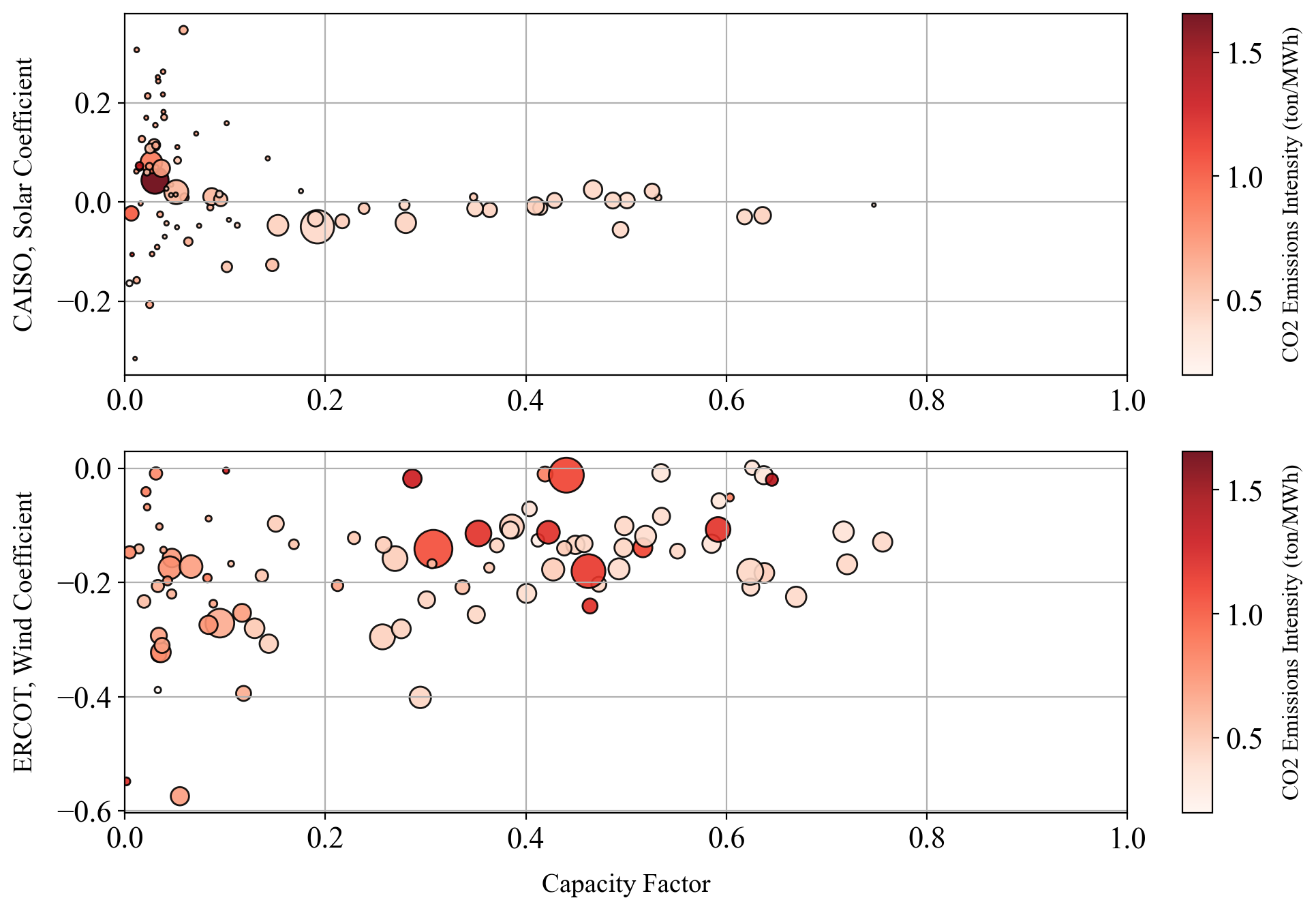}
    \caption{Correlation between annual plant capacity factor in 2022 and the solar coefficient for generation from natural gas plants in CAISO (above) and the wind coefficient for coal and natural gas plants in ERCOT (wind). The size of the points indicates the nameplate capacity while the shaded color represents the annual emissions intensity in 2022.}
    \label{fig:coeff-cf_ei}
\end{figure*}

\subsection{Individual plant response in CAISO and ERCOT is heterogeneous by fuel type and nameplate capacity}

The panel regression formulation above helps us understand how thermal plants across the fleet respond to changes in generation from wind and solar. Location-based fixed-effects further control for heterogeneity across plant types and fuel inputs, but this formulation does not yield any insights into how heterogeneous these plants are, and what the impact of increased generation from wind and solar results in on a per-plant basis. To further understand plant-specific response, we develop a fixed-effects ordinary least squares (OLS) regression formulation with the same specification as the panel regression above, except for every individual plant (no panel is considered) and using hourly instead of daily data points. The ramp variable in the formulation is determined as a difference of wind generation at the intra-hourly level. While the panel regression provided a single coefficient for a dependent variable in response to solar and wind, here, we consider a distribution of coefficients across all plants to model their response.

Figure \ref{fig:map-coeff} shows the distribution of coefficients for plant generation in response to solar in CAISO (left) and wind in ERCOT (right). A positive coefficient implies that generation increases in response to generation from renewables and vice versa. We find that natural gas plants in CAISO show a very heterogeneous response to generation from solar with plants responding via an increase as well as a decrease in generation. In contrast, all but two plants in ERCOT lower their generation in response to increasing generation from wind. 

The heterogeneous response of natural gas plants in CAISO may be attributed to a shift in generation patterns as more utility-scale solar comes online. Fast-ramping peaker plants replace base load combined cycle plants during peak production hours. In CAISO, 4 out of 7 plants with a coefficient greater than 1.0 account for 28\% of generation and also exhibit the highest emissions intensity of all plants in the balancing region. In ERCOT, plant behavior is very different to that of CAISO. The mean coefficient for generation is -0.41, which is close to our fleet level coefficient of -0.34 from the panel regression. Coal plants with a relatively smaller nameplate capacity are characterized by a higher emissions intensity, even though the capacity factor of these plants is comparable to that of larger plants. The two smallest coal plants in ERCOT show the largest deviation in response to wind (-2.25\%, -2.07\%) and also have a $\text{NO}_x$ emissions intensity 60\% higher than the mean of all coal plants indicating more frequent ramping and startup.

Figure \ref{fig:coeff-cf_ei} shows the regression coefficients for plant generation. Each point represents a plant. The x axis is annual capacity factor, and the y axis is the OLS coefficient in response to solar in CAISO and wind in ERCOT. Points are sized by plant nameplate capacity and colored by their annual emissions intensity. In CAISO, all of the plants are natural gas-fired. We find that the larger plants lower their generation in response to increased generation from solar, although the magnitude of this deviation is smaller compared to plants with a lower nameplate capacity. Smaller plants are also characterized by a significantly lower annual capacity factor, high emissions intensity, and a wider distribution in coefficient magnitude.

In ERCOT, the points in Figure \ref{fig:coeff-cf_ei} show a mix of coal and natural gas-fired plants. Coal plants have a higher emissions intensity, and also operate at a annual capacity factor consistently greater than 0.3. The magnitude of their response with increase generation from wind is relatively less, with larger plants not deviating beyond -0.2. Similarly to CAISO, natural gas plants with a lower nameplate capacity show the highest emissions intensity and operate at lower annual capacity factors.

Figures \ref{fig:kde_CAISO} and \ref{fig:kde_ERCOT} in Section \ref{comparison} in the SI show a comparison between the distrubution of coefficients obtained from the plant-level OLS with emissions displacement estimates from Katzenstein and Apt \cite{katzenstein2009air} and Graf et al. \cite{graf2017renewable}.

\section{Discussion}

In this study, we analyze the impact of accelerated grid-integration of utility-scale wind and solar, and the resulting change in operational characteristics of coal and natural gas plants in two regions in the Unites States - CAISO and ERCOT. Using hourly power plant generation and emissions data obtained from CEMS, we find that operating power plants at low capacity factors significantly affects their heat rate and emissions intensity. We use three scenarios to model the effect of running thermal plants at various points in their operational envelope. With this empirical method, we see that there is scope to further lower system emissions if plants are operated above a certain capacity factor threshold. We also highlight how constant emissions intensity values used in capacity expansion and unit commitment models can severely under-represent gross emissions especially in high-renewables scenarios of the electric grid.

Further, we use a daily fixed effects panel regression model with a logarithmic specification to attribute the impact of renewables on thermal plants in CAISO and ERCOT. At a systems-level, we find that renewables displace thermal generation and emissions in both regions, although the magnitude of the displacement varies by the renewable resource. In contrast, the emissions intensity for all plants increases, albeit marginally compared to the reduction in generation and emissions. If the displacement of emissions due to renewables was 100\% efficient, the coefficients of the panel regression with generation and emissions as the dependent variable would be identical. For instance, if thermal generation for a plant reduces by 0.2\% in response to solar with no impact on the heat rate or emissions intensity, then we would expect the emissions to also reduce by the same magnitude. We find that two coefficients are not identical, with the coefficient for emissions always lagging behind that of generation for both wind and solar. Using the logarithmic specification for emissions and emissions intensity, we thus quantify the share of expected displacement by wind and solar on thermal plants.

In addition to the panel regression model that describes plant response for the fleet, we use an OLS formulation with a similar structure except applied on hourly data points and modeled for individual thermal plants. We find that plant response is very heterogeneous in both regions and also varies by the displacing renewable resource. Overall, plants with a smaller nameplate capacity show a wider envelope of solar and wind coefficients in response to increasing renewable generation, and are also characterized by a lower annual capacity factor and high emissions intensity. 

In summary, we show that the first order effect of increasing generation from renewables is lowered emissions and generation from thermal power plants, however the displacement is not 1:1 as expected. In fact, smaller peaker plants are more prone to ramping up and down at a lower capacity factor and high heat rate, resulting in higher instantaneous emissions. As more renewables are installed in the grid, we expect that traditionally baseload plants will be increasingly made to cycle in a manner similar to the peaker plants, leading to a further inefficiency at the plant level. If system operators are to achieve a combined thermal displacement and emissions reduction goal, it is necessary for these secondary impacts to be addressed by appropriate intensity limits and regulation of spinning reserves. 

\section{Data and Methods}

We perform our analysis for years 2018 through 2023 for the California Independent System Operator (CAISO) and the Electricity Reliability Council of Texas (ERCOT). We start by describing the data used, followed by the methods.

\subsection{Data}

We develop a dataset with  thermal power plants in CAISO and ERCOT that includes hourly unit-level emissions, gross generation, heat rate, and fuel consumption from the Environmental Protection Agency’s (EPA) Air Markets Program Data (AMPD) \cite{CEMS}. The EPA requires all combustion power plants with a nameplate capacity greater than 25 MW to install and maintain a continuous emission monitoring system (CEMS) which records key unit-level operational parameters \cite{us2009plain}.

We obtain plant nameplate capacity, total annual generation, primary fuel information, and total annual emissions from the EPA's Emissions and Generation Resource Integrated Database (eGRID) \cite{EGRID}. eGRID is a comprehensive source of data on the environmental characteristics of electric power plants and their associated unit stacks. Each data file contains key indicators at different levels of aggregation - unit, generator, plant, state, balancing area, NERC region and eGRID subregion. A manual inspection of the plant-level nameplate capacity indicates that the value indicated is not accurate in all cases, whereas the same is not true for aggregated generator-level information. Hence, instead of obtaining the plant nameplate capacity directly, we compute the sum of the generator nameplate capacity (in MW) of all generators associated with a particular plant ORISPL. Hourly data from CEMS is collated with plant nameplate capacity to determine hourly capacity factor values and the associated fuel stack.

Hourly generation from wind, solar, hydro, and geothermal energy is obtained from US Energy Information Administration Form 930 (EIA-930) data. We use processed publicly available data obtained from a physics informed data reconciliation framework for real-time electricity and emissions tracking developed by Chalendar and Benson \cite{de2019tracking,de2020estimating}. Using this dataset, rather than raw EIA-930 data, has two key advantages. First, the reconciled data sets are devoid of unrealistic generation values, and contain estimates within reasonable confidence intervals for time intervals where data is missing or erroneous. Second, Chalendar and Benson perform an optimization-based data reconciliation to impose physical relations that guarantee energy conservation between key variables. Thus, this dataset provides electric system operating data on generation, consumption, and exchanges of electricity for every hour and at the level of the balancing area.

\subsection{Methods}

\subsubsection{Statistical models}

We use a fixed-effects panel regression formulation to understand the marginal effect of wind and solar generation on different dependent variables of interest, namely: generation, emissions, and emissions intensity of thermal plants. Similar to Bushnell and Wolfram \cite{bushnell2005ownership},  Graf et al. \cite{graf2017renewable} and Fell and Johnson \cite{fell2021regional}, we use a fixed effects logarithmic specification for all plants within an ISO using daily observations. Our approach departs from those previous models in that we explicitly control for the total generation of therm power plans withing the ISO, total hydro generation within the ISo, net imports, and the generation from wind and from solar. We further also account for wind variability. Our model specification is as follows:

\begin{multline}
     \ln{y_{i,t}} = \alpha_i + \beta_1 \ln{G_t} + \beta_2 \ln{S_t} + \beta_3 \ln{W_t} + \beta_4 \ln{\overline{W}_t} \\ 
     + \gamma \ln{\textbf{X}_t} + \omega \ln{\textbf{Y}_t} + \eta_{t,m} + \eta_{t,a} + \alpha_i \times \eta_{t,m} + \epsilon_i
\label{log_spec_1}
\end{multline}

\noindent where $y_{i,t}$ is the dependent variable of interest, which is either the daily generation in MWh, emissions in tonnes or the emissions intensity in tonnes/MWh for power plant $i$ at timestep $t$. $G_t$, $S_t$ and $W_t$ represent the daily total thermal generation, generation from solar, and generation from wind in MWh for the ISO to which $i$ belongs. $\overline{W}_t$ is a parameter that estimates the intermittency of net generation from wind at the hourly level and is calculated by taking the sum of the absolute value of hourly differences in wind output over the course of the day. $\textbf{X}_t$ is a set of control variables that includes the sum of wind generation, solar generation, and demand for trading partners of the ISO to which $i$ belongs. For instance, CAISO trades with the other balancing authorities in the Northwest (NW) and Southwest (SW) NERC subregions. For every hour, this external control variables would consider the sum of wind generation, solar generation, and demand for these two regions. For ERCOT, the only trading partner is the Southwest Power Pool (SWPP). $\textbf{Y}_t$ is a set of control variables that include the hydro and imports for the ISO. $\eta_{t,m}$ and $\eta_{t,a}$ are the time fixed-effects parameters controlling for the month and year, respectively. We interact the monthly fixed-effects variable with the indicator variable for the plant ID to account for heterogeneity in fuel prices across different technology categories, regions and plant types. $\epsilon$ represents the error term.

The model specifications employed by Graf et al. \cite{graf2017renewable} and Fell and Johnson \cite{fell2021regional} may not be able to fully explain the deviation caused due to intermittency in renewable generation due to gaps in granularity and panel formulation. Graf et al. \cite{graf2017renewable}do not include granular regional generation from trading partners, and also conduct the analysis on annual data and hence miss intra-day operational parameters such as resource ramping. Fell and Johnson \cite{fell2021regional} do not construct a plant-level panel regression but rather run a time fixed effects model that does not account for location-based heterogeneity.

In our formulation, the parameters that are of interest are the coefficients for solar, $\beta_2$ and wind, $\beta_3$. These coefficients represent the percent change in the dependent variable per 1\% increase in generation from solar and wind, respectively. For example, if the dependent variable $y_{i,t}$ represents the net generation from thermal plants for an ISO, then $\beta_2$ and $\beta_3$ will represent the percent change in generation for thermal plants per unit percent increase in generation from solar and wind. These coefficients thus measure the causal change in the dependent variable under the assumption that daily wind and solar production (represented by $W$ and $S$) is uncorrelated with the error term $\epsilon$ after controlling for net demand, entity, and time fixed effects. As suggested by \cite{qiu2022impacts}, this assumption is valid for wind at the hourly level and thus can be extrapolated to daily observations. Solar output is likely to be correlated with the error term and time fixed effects parameter when considering hourly data, which is why we aggregate hourly observations to daily intervals in the statistical model above.

We also present alternate specifications of the statistical model in Section \ref{alternate} in the SI.

\subsubsection{Plant-level ordinary least squares regression}

Modeling the response of individual plants outside a panel regression formulation helps us understand whether plants are individually load-following, solar-following, and wind-following, and the extent of their response. In contrast to fixed effects ISO-level formulation considering daily time steps, the plant-level model uses hourly data and is given by:

\begin{multline}
     \ln{y_{t}} = \alpha + \beta_1 \ln{G_t} + \beta_2 \ln{S_t} + \beta_3 \ln{W_t} + \beta_4 \ln{\overline{W}_t} \\ 
     + \gamma \ln{\textbf{X}_t} + \omega \ln{\textbf{Y}_t} + \eta_{t,m} + \eta_{t,a} + \epsilon
\label{log_spec_ols}
\end{multline}

\noindent where the variables are the same as those in the panel regression formulation above.

\subsubsection{Deviation between observed and actual emissions}

\label{deviation}

To determine the share of emissions displaced due to renewables, we use the panel regression formulations where emissions and emissions intensity are the dependent variables similar to Graf et al. \cite{graf2017renewable}.

\begin{multline}
     \ln{\text{CO}_2\text{EI}_{p,t}} = \alpha_i + \beta_1 \ln{G_t} + \beta_2 \ln{S_t} + \beta_3 \ln{W_t} + \beta_4 \ln{\overline{W}_t} \\ 
     + \gamma \ln{\textbf{X}_t} + \omega \ln{\textbf{Y}_t} + \eta_{t,m} + \eta_{t,a} + \alpha_i \times \eta_{t,m} + \epsilon_i
\label{log_spec_co2ei}
\end{multline}

\begin{multline}
     \ln{\text{CO}_{2,p,t}} = \alpha_i’ + \beta_1’ \ln{G_t} + \beta_2 ‘\ln{S_t} + \beta_3’ \ln{W_t} + \beta_4’ \ln{\overline{W}_t} \\ 
     + \gamma’ \ln{\textbf{X}_t} + \omega \ln{\textbf{Y}_t} + \eta’_{t,m} + \eta’_{t,a} + \alpha_i’ \times \eta’_{t,m} + \epsilon’_i
\label{log_spec_co2}
\end{multline}

In Equation \ref{log_spec_co2ei}, the marginal effect due to increased generation from solar can be represented by:

\begin{equation}
    \beta_2 = \frac{\partial(\ln{\text{CO}_2\text{EI}_{p,t}})}{\partial(\ln{\text{solar}_{t}})}
\end{equation}

\begin{equation}
    \beta_2 = \frac{\partial(\ln{\frac{\text{CO}_{2,p,t}}{G_{p,t}}})}{\partial(\ln{\text{solar}_{t}})}
\end{equation}

\begin{equation}
    \beta_2 = \frac{\partial(\ln\text{CO}_{2,p,t} - \ln{G_{p,t}})}{\partial(\ln{\text{solar}_{t}})}
\end{equation}

\begin{equation}
    \beta_2 = \frac{\partial(\ln\text{CO}_{2,p,t})}{\partial(\ln{\text{solar}_{t}})} - \frac{\partial(\ln G_{p,t})}{\partial(\ln{\text{solar}_{t}})}
\end{equation}

\begin{equation}
    \beta_2 = \beta_2' - \frac{\partial(\ln G_{p,t})}{\partial(\ln{\text{solar}_{t}})}
\end{equation}

Rearranging, gives:

\begin{equation}
    \frac{\partial(\ln G_{p,t})}{\partial(\ln{\text{solar}_{t}})} = \beta_2 - \beta_2' 
\end{equation}

The final equation represents the change in thermal generation if generation from solar replaces the generation from fossil-fueled plants on a 1:1 basis. Thus, the fraction of expected emissions reductions for solar is represented by:

\begin{equation}
    \text{Fraction of expected emissions reduction (solar)} = \frac{\beta_2'}{\beta_2'-
    \beta_2} 
\end{equation}

Similarly, for wind, the expected emissions reductions and the corresponding fraction are given by:

\begin{equation}
    \frac{\partial(\ln G_{p,t})}{\partial(\ln{\text{wind}_{t}})} = \beta_3 - \beta_3' 
\end{equation}

\begin{equation}
    \text{Fraction of expected emissions reduction (wind)} = \frac{\beta_3'}{\beta_3'-
    \beta_3} 
\end{equation}

\subsubsection{Determining generation ramp rate}

Plant ramp rate is determined by using two different approaches that enable us to understand operational behavior at different temporal resolutions. Analyzing daily variation in maximum and minimum utilization for a plant helps us understand the fluctuation in thermal plant generation over the course of 24 hours, whereas intra-day ramping measured in MW change in output per hour provides us with more granular insights on how fast or slow certain plant types are responding to variable generation from wind and solar. We follow definitions for ramping presented in Triolo and Wolak \cite{triolo2022quantifying} definitions, where daily ramping is computed as the maximum hourly output – minimum hourly output on a given day. Intra-day hourly ramping is determined by calculating the difference between the net generation at time $t$ and the previous time interval $t-1$.

\bibliographystyle{ieeetr}
\bibliography{bibliography}

\newpage

\appendices

\onecolumn

\section{}

\subsection{Emissions intensity as a function of capacity factor}
\label{EI_CF_plots}

\begin{figure}
    \centering
    \includegraphics[width=\linewidth]{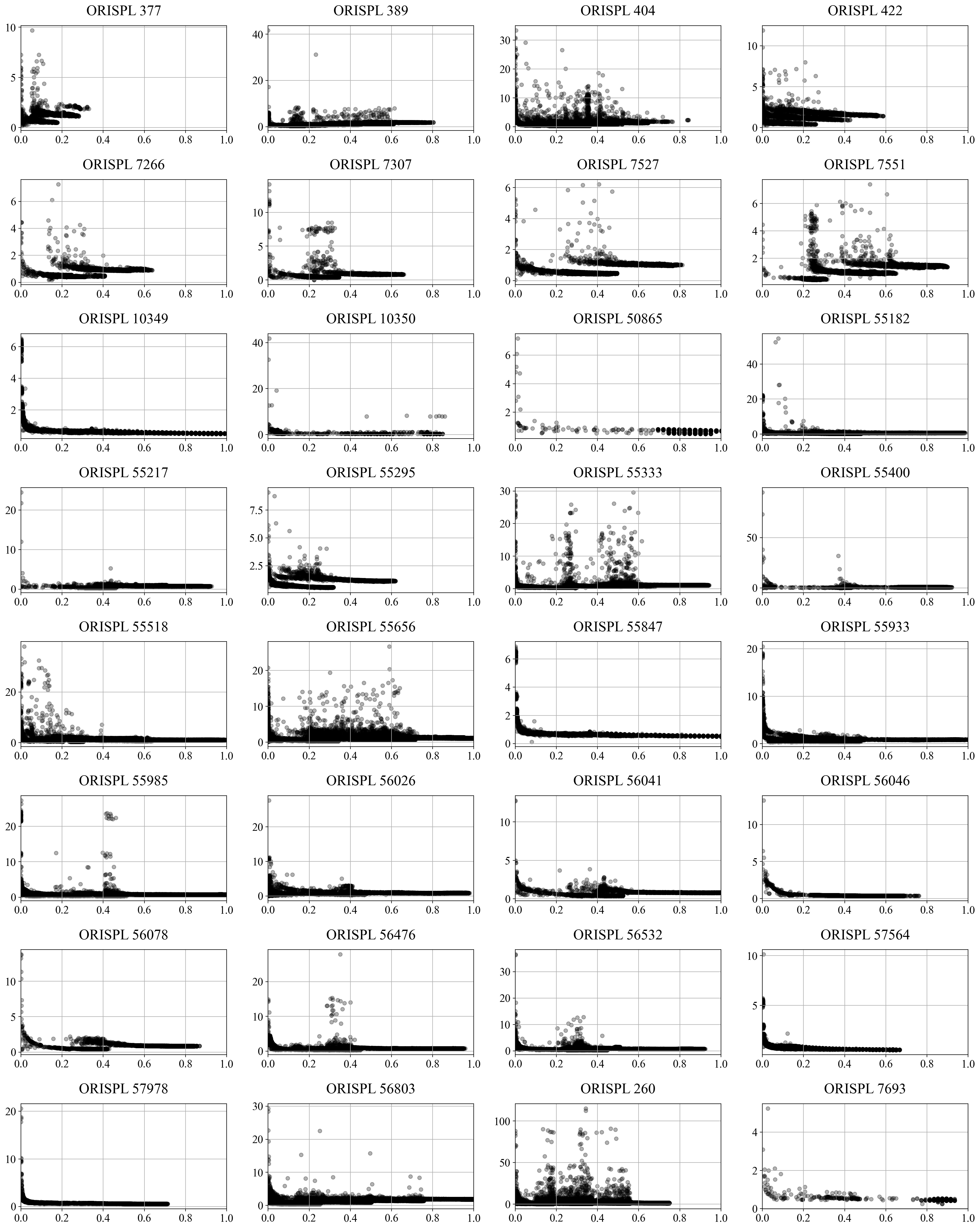}
    \caption{Relation between emissions intensity and capacity factor for Natural Gas plants in CAISO}
    \label{fig:hourly_metrics-2}
\end{figure}

\begin{figure}
    \centering
    \includegraphics[width=\linewidth]{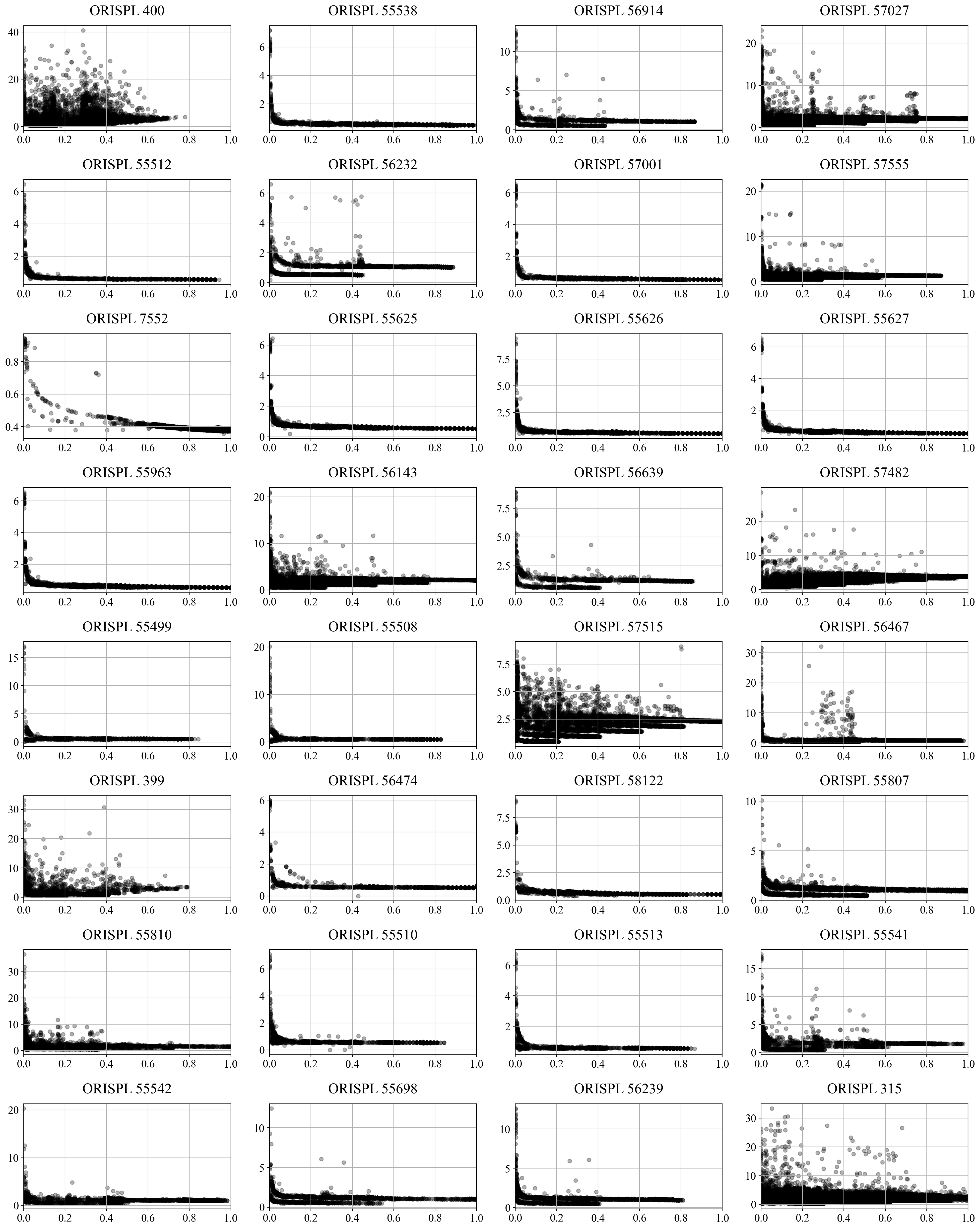}
    \caption{Relation between emissions intensity and capacity factor for Natural Gas plants in CAISO}
    \label{fig:hourly_metrics-3}
\end{figure}

\begin{figure}
    \centering
    \includegraphics[width=\linewidth]{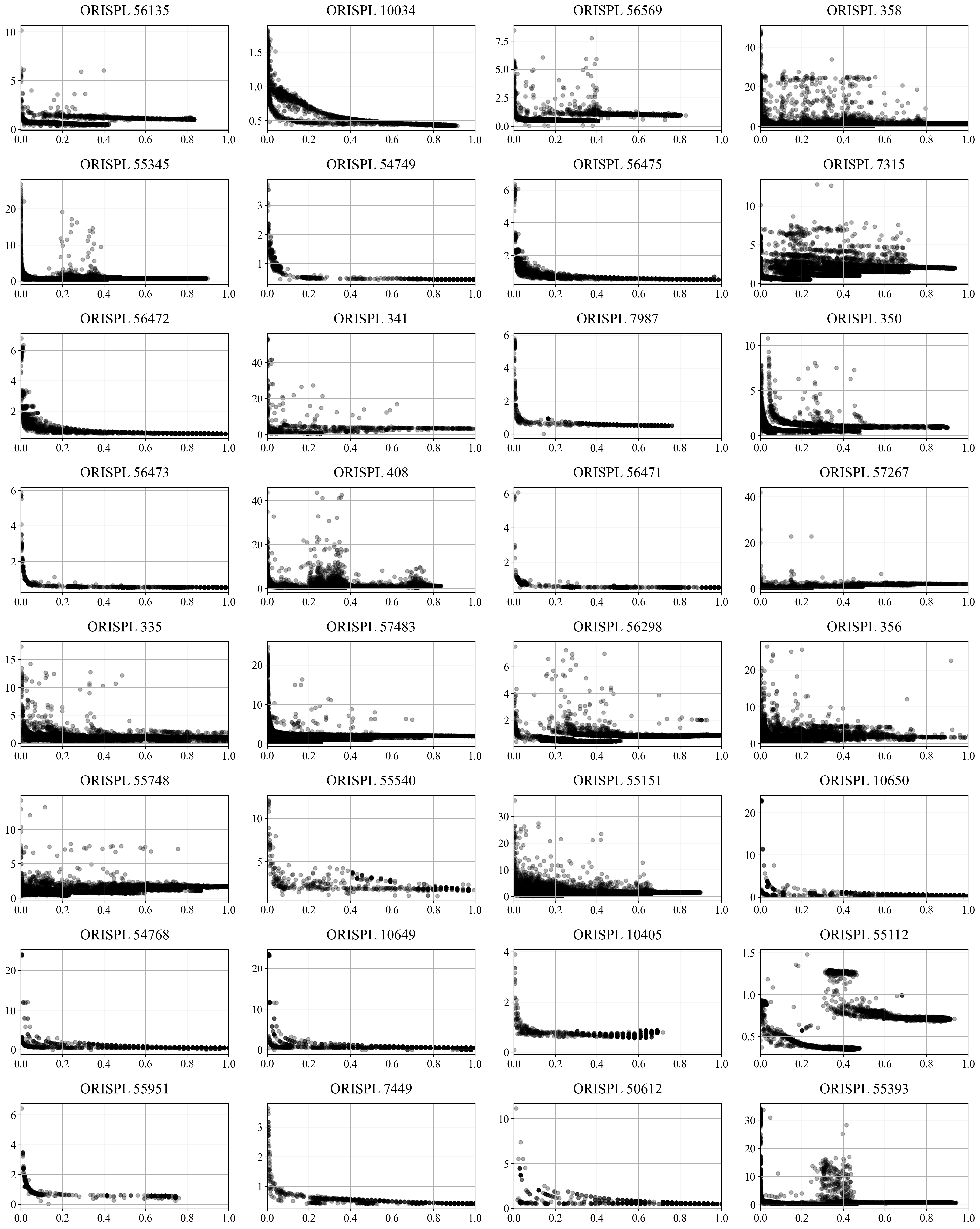}
    \caption{Relation between emissions intensity and capacity factor for Natural Gas plants in CAISO}
    \label{fig:hourly_metrics-4}
\end{figure}

\begin{figure}
    \centering
    \includegraphics[width=\linewidth]{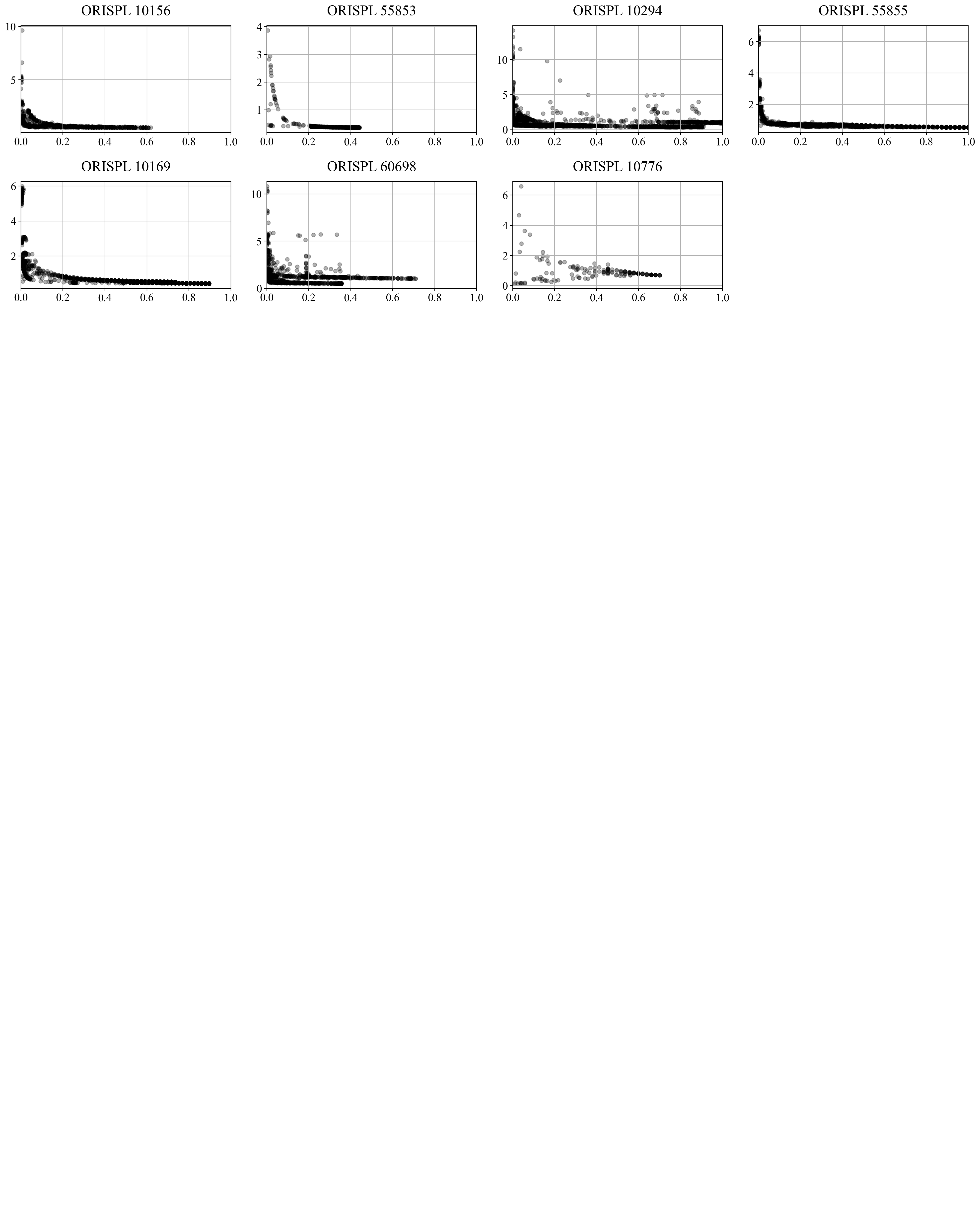}
    \caption{Relation between emissions intensity and capacity factor for Natural Gas plants in CAISO}
    \label{fig:hourly_metrics-5}
\end{figure}

\begin{figure}
    \centering
    \includegraphics[width=\linewidth]{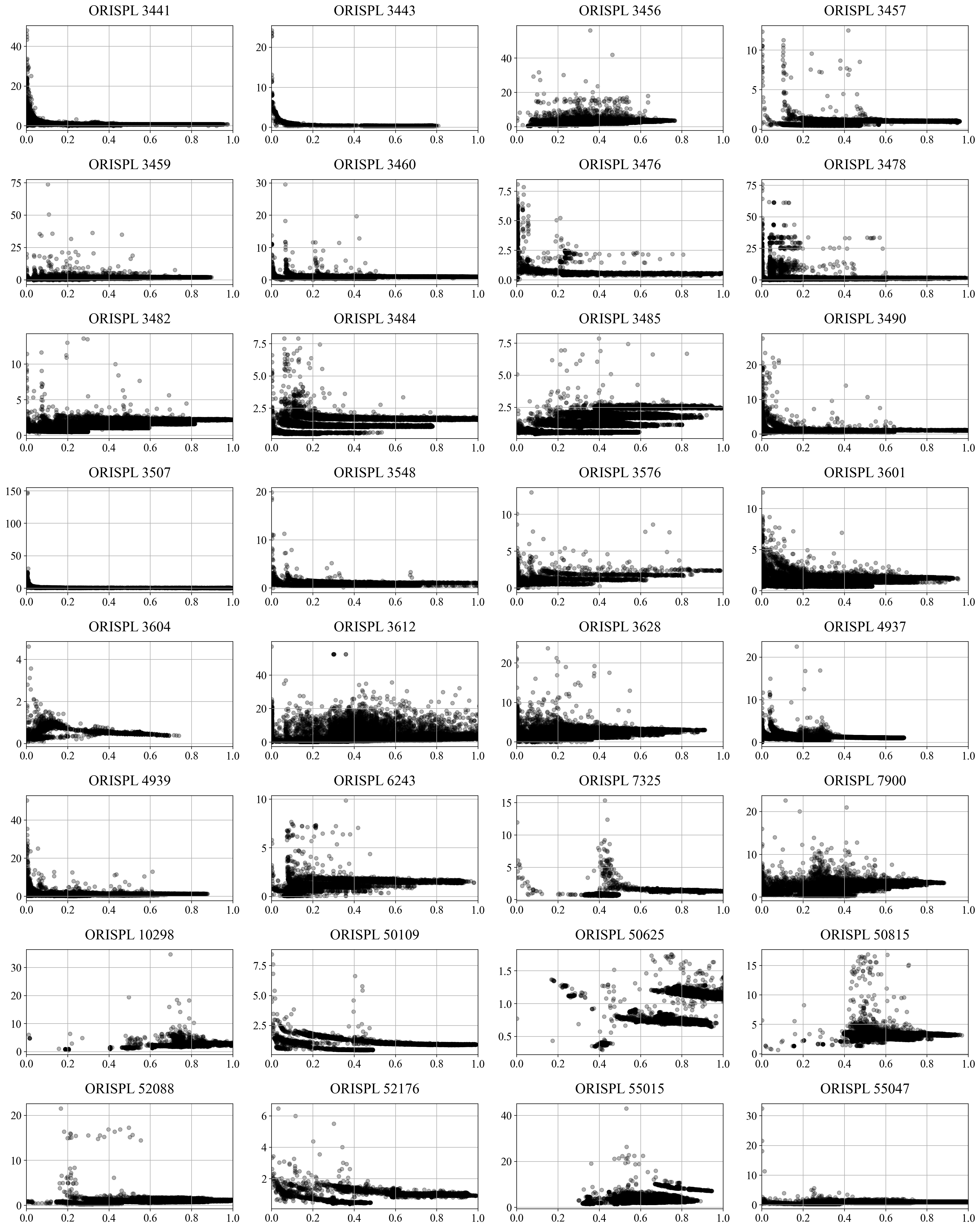}
    \caption{Relation between emissions intensity and capacity factor for Natural Gas plants in ERCOT}
    \label{fig:hourly_metrics-6}
\end{figure}

\begin{figure}
    \centering
    \includegraphics[width=\linewidth]{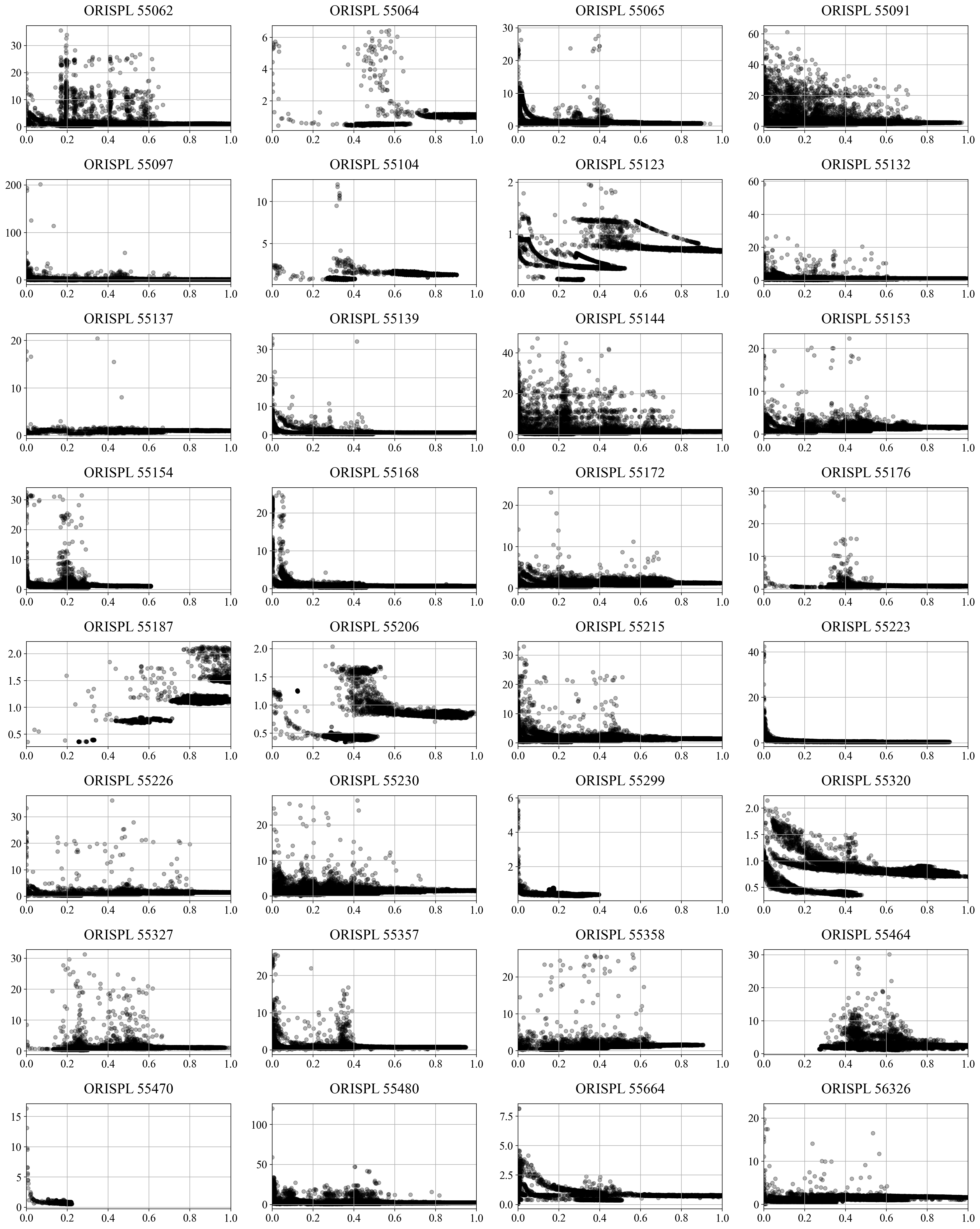}
    \caption{Relation between emissions intensity and capacity factor for Natural Gas plants in ERCOT}
    \label{fig:hourly_metrics-7}
\end{figure}

\begin{figure}
    \centering
    \includegraphics[width=\linewidth]{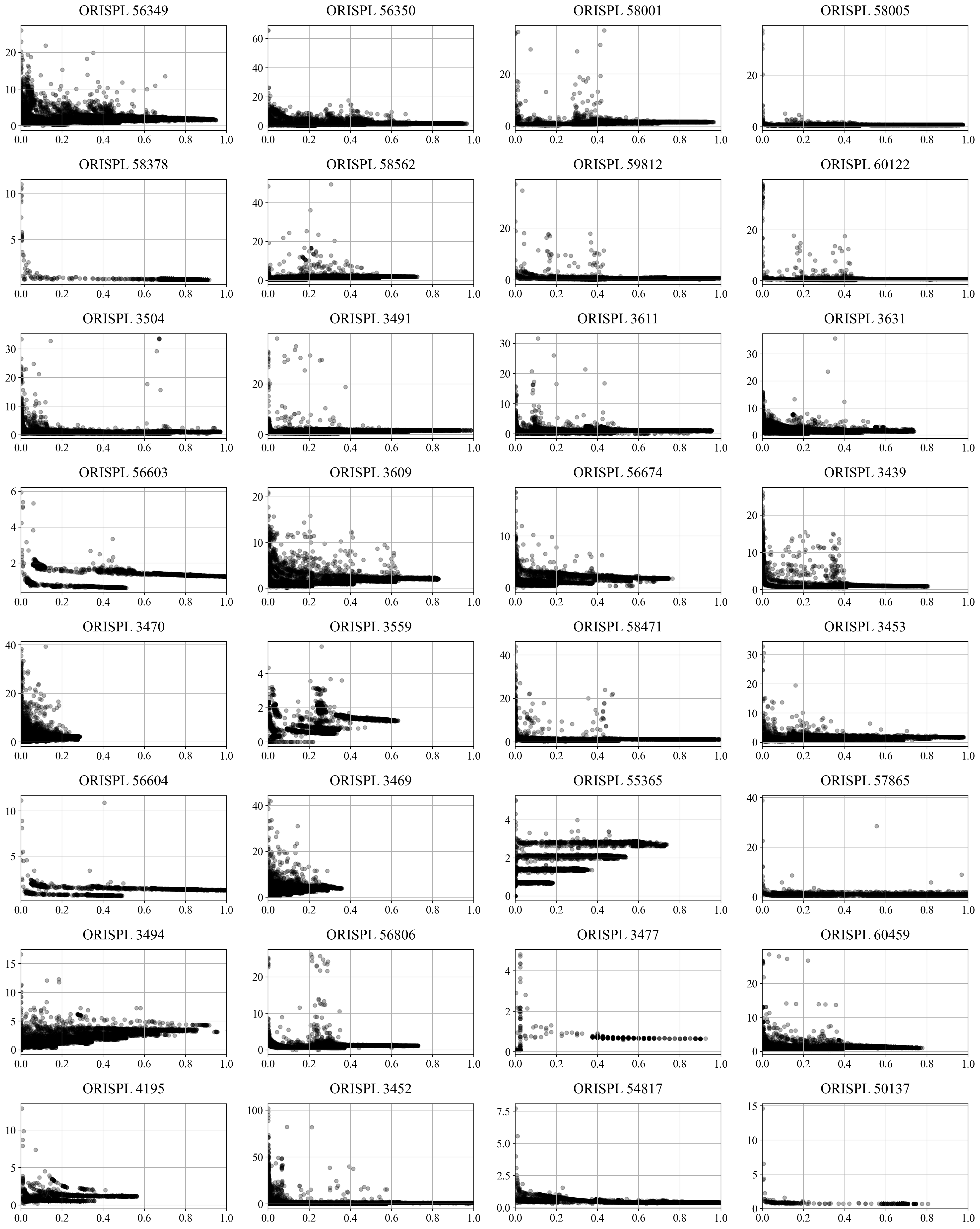}
    \caption{Relation between emissions intensity and capacity factor for Natural Gas plants in ERCOT}
    \label{fig:hourly_metrics-8}
\end{figure}

\begin{figure}
    \centering
    \includegraphics[width=\linewidth]{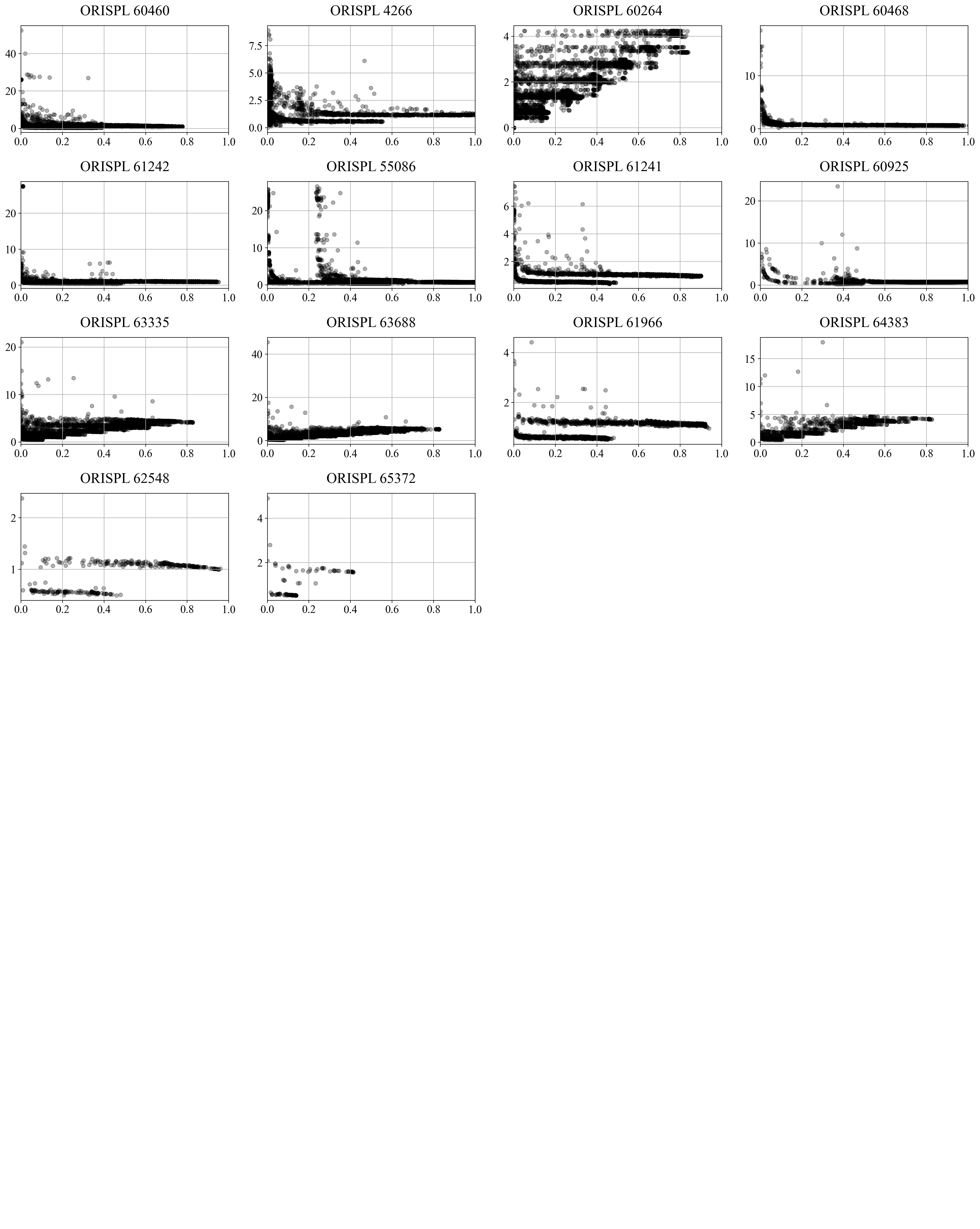}
    \caption{Relation between emissions intensity and capacity factor for Natural Gas plants in ERCOT}
    \label{fig:hourly_metrics-9}
\end{figure}

\begin{figure}
    \centering
    \includegraphics[width=\linewidth]{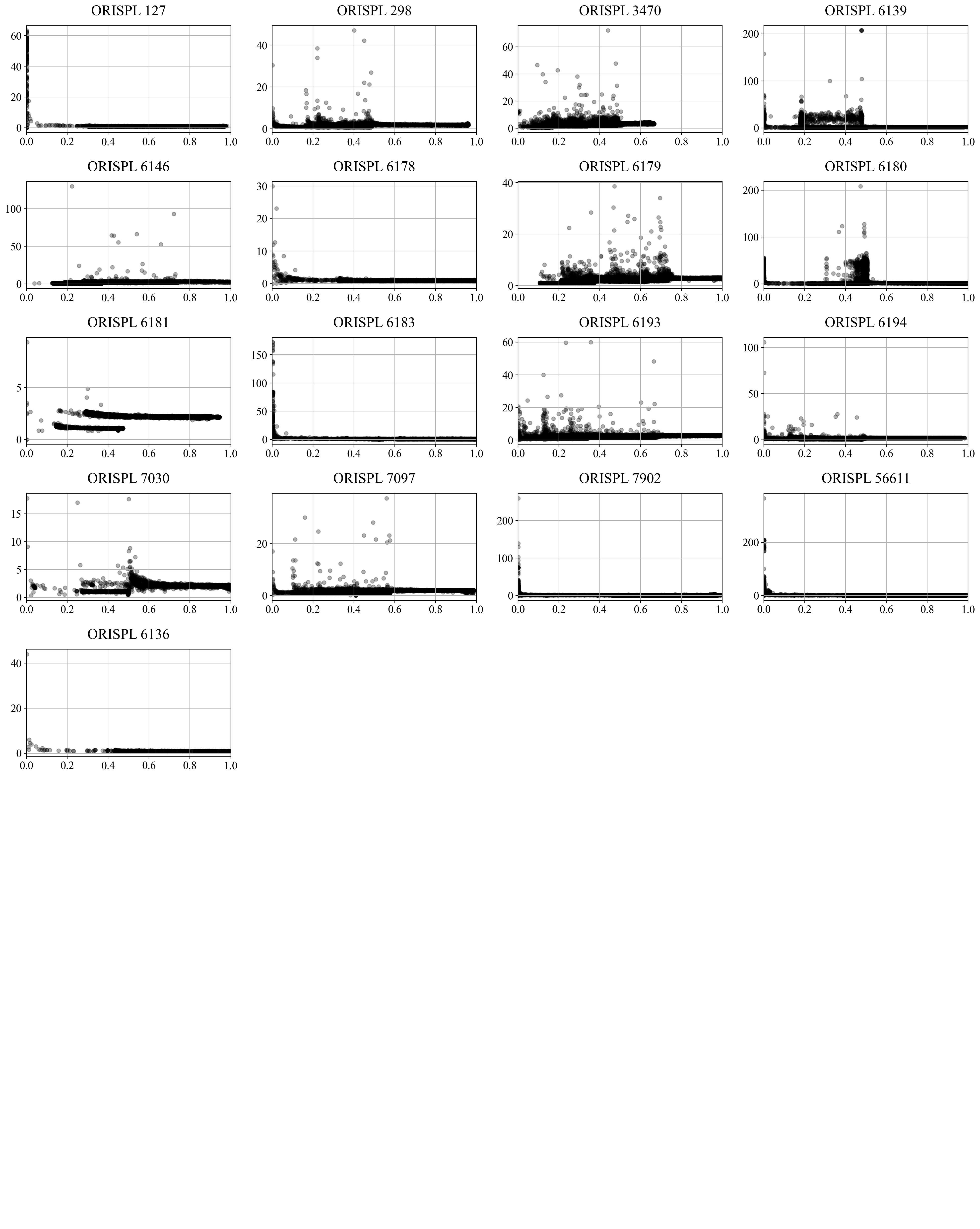}
    \caption{Relation between emissions intensity and capacity factor for Coal plants in ERCOT}
    \label{fig:hourly_metrics-10}
\end{figure}

Figures \ref{fig:hourly_metrics-2} through \ref{fig:hourly_metrics-10} show the variation of emissions intensity as a function of capacity factor for thermal plants in CAISO and ERCOT.

\newpage
\subsection{Change in operational behaviour for thermal plants}
\label{change_in_op}

\begin{figure}
    \centering
    \includegraphics[width=\linewidth]{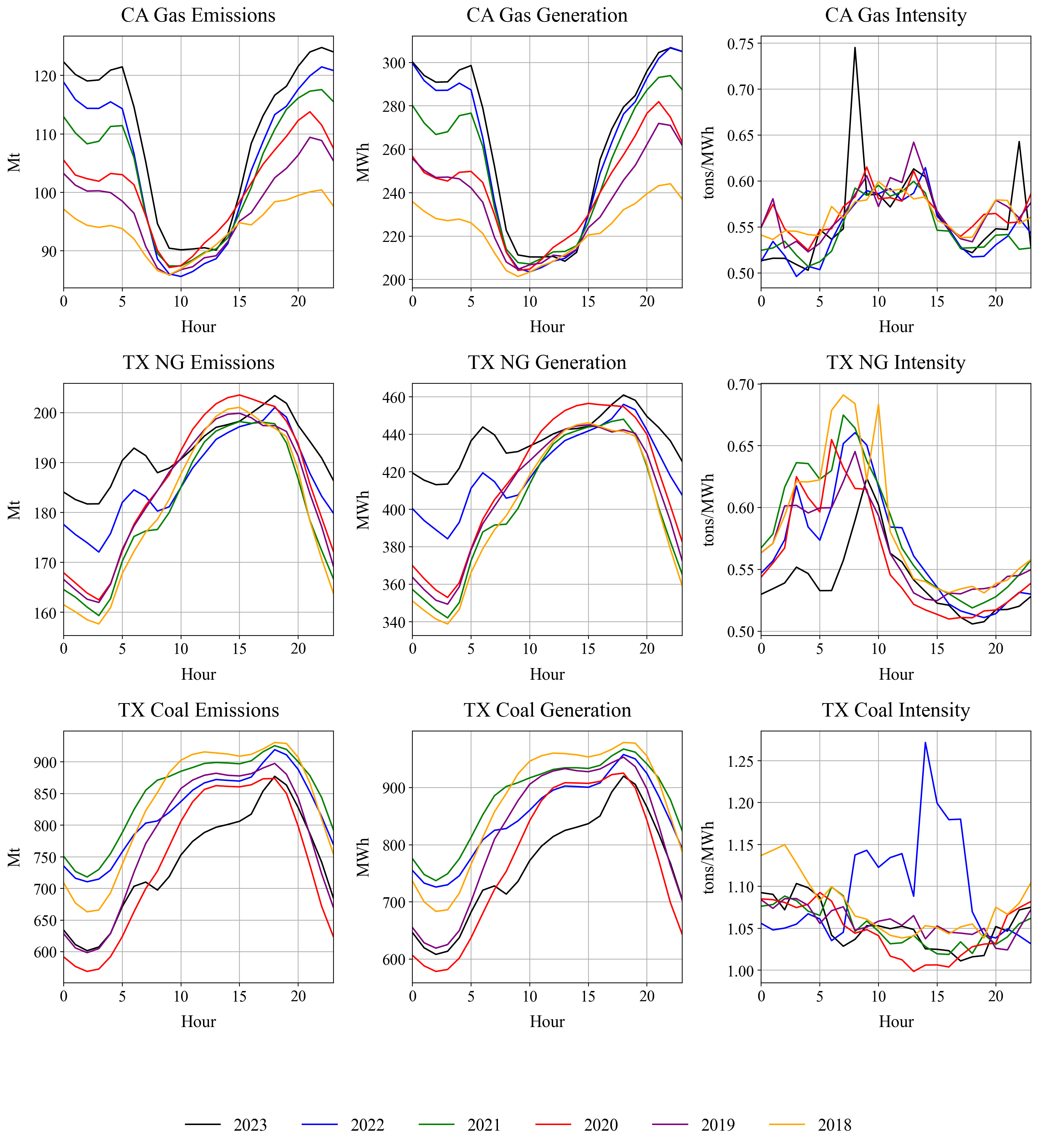}
    \caption{Aggregated emissions, generation, and emissions intensity for (top) natural gas plants in CAISO, (middle) natural gas plants in ERCOT, and (bottom) coal plants in ERCOT.}
    \label{fig:hourly_metrics-1}
\end{figure}

Figure \ref{fig:hourly_metrics-1} shows the mean hourly emissions, generation and emissions intensity across all plants types in CAISO and ERCOT. The x axis shows the hour of day, while the y axis measures the key variable of interest.

\subsection{Comparison with previous studies}
\label{comparison}

\begin{figure*}[ht]
    \centering
    \begin{subfigure}[b]{0.6\linewidth}
        \includegraphics[width=\linewidth]{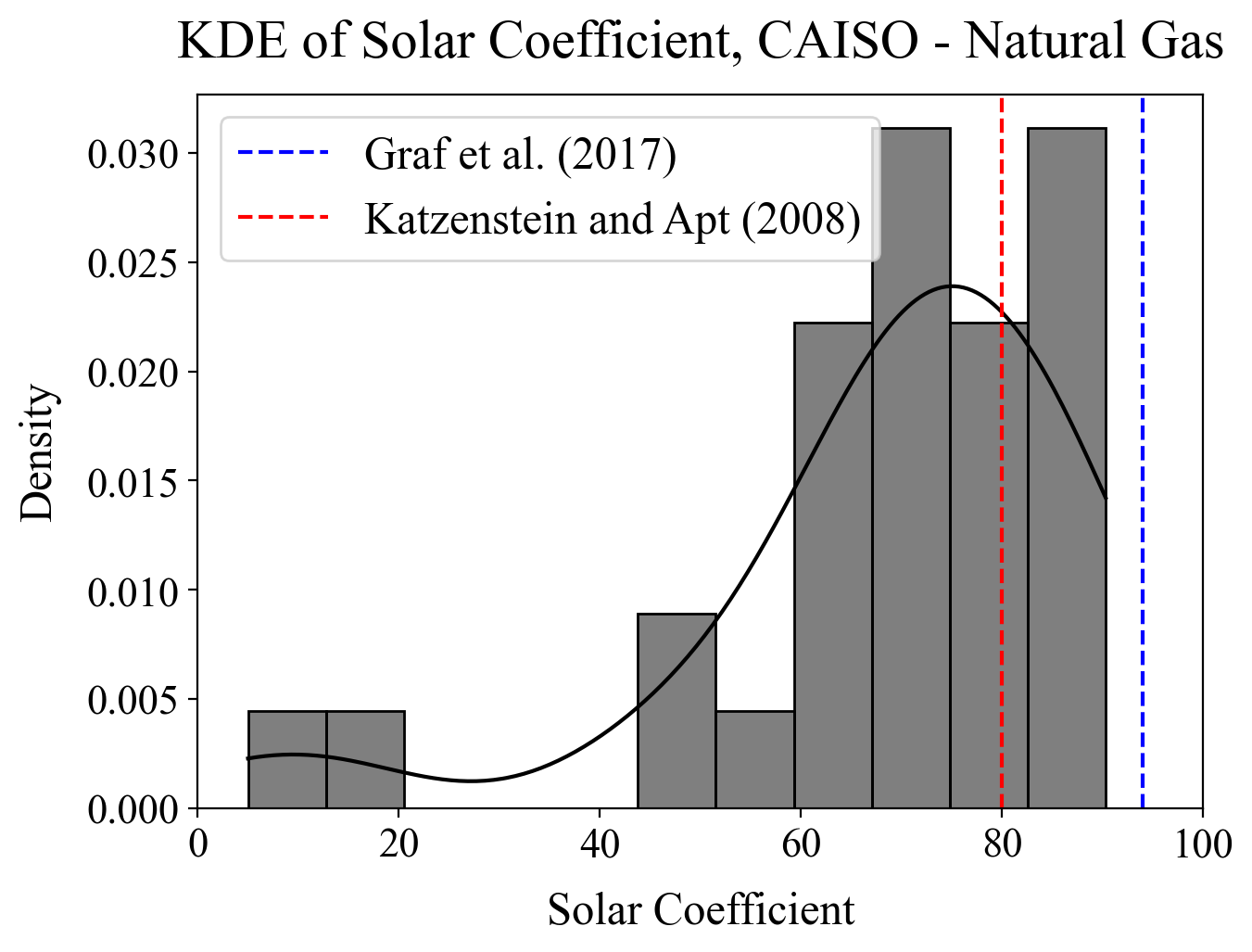}
        \caption{KDE and histogram of individual plant coefficients in response to solar generation for natural gas plants in CAISO.}
        \label{fig:kde_CAISO}
    \end{subfigure}
    \hfill
    \begin{subfigure}[b]{0.6\linewidth}
        \includegraphics[width=\linewidth]{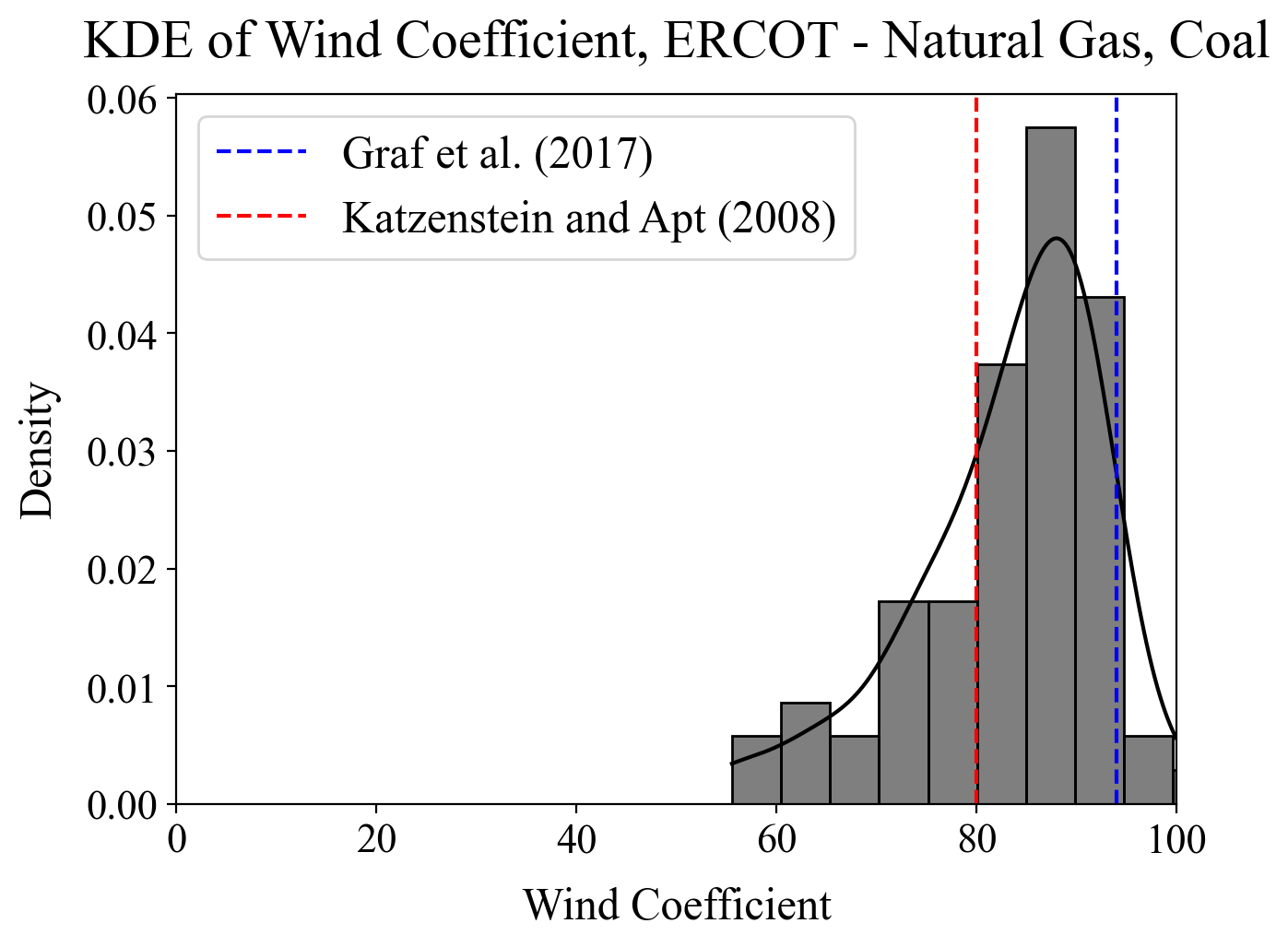}
        \caption{KDE and histogram of individual plant coefficients in response to wind generation for coal and natural gas plants in ERCOT.}
        \label{fig:kde_ERCOT}
    \end{subfigure}
    \caption{Comparative analysis of emissions displacement due to renewable energy sources in CAISO and ERCOT. The red and blue dashed lines indicate the expected vs actual emissions displacement computed by Graf et al. and Katzenstein and Apt.}
\end{figure*}

In Figures \ref{fig:kde_CAISO} and \ref{fig:kde_ERCOT}, we show the distribution of expected vs actual emissions reductions calculated through the plant OLS model. We compare these individual coefficient estimates to the fleet-level coefficient from the panel regression, as well as to Graf et al. and Katzenstein and Apt's analysis.

\subsection{Alternate specifications of the statistical model}
\label{alternate}

We test several specifications of the location and time fixed effects model. Model coefficients, standard errors, R-squared values and key parameters are summarized in Table \ref{table:coefficients-CAISO} and \ref{table:coefficients-ERCOT} below.

\noindent Main specification

\begin{flalign}
    &\textbf{M1:} \hspace{1cm}
    \ln{y_{i,t}} = \alpha_i + \beta_1 \ln{G_t} + \beta_2 \ln{S_t} + \beta_3 \ln{W_t} + \beta_4 \ln{\overline{W}_t}
     + \gamma \ln{\textbf{X}_t} + \omega \ln{\textbf{Y}_t} + \eta_{t,m} + \eta_{t,a} + \alpha_i \times \eta_{t,m} + \epsilon_i &
\end{flalign}

\noindent Exclude demand

\begin{flalign}
    &\textbf{M2:} \hspace{1cm}
    \ln{y_{i,t}} = \alpha_i + \beta_2 \ln{S_t} + \beta_3 \ln{W_t} + \beta_4 \ln{\overline{W}_t}
     + \gamma \ln{\textbf{X}_t} + \omega \ln{\textbf{Y}_t} + \eta_{t,m} + \eta_{t,a} + \alpha_i \times \eta_{t,m} + \epsilon_i &
\end{flalign}

\noindent Exclude solar generation

\begin{flalign}
    &\textbf{M3:} \hspace{1cm}
    \ln{y_{i,t}} = \alpha_i + \beta_1 \ln{G_t} + \beta_3 \ln{W_t} + \beta_4 \ln{\overline{W}_t}
     + \gamma \ln{\textbf{X}_t} + \omega \ln{\textbf{Y}_t} + \eta_{t,m} + \eta_{t,a} + \alpha_i \times \eta_{t,m} + \epsilon_i &
\end{flalign}

\noindent Exclude wind generation

\begin{flalign}
    &\textbf{M4:} \hspace{1cm}
    \ln{y_{i,t}} = \alpha_i + \beta_1 \ln{G_t} + \beta_2 \ln{S_t} + \beta_4 \ln{\overline{W}_t}
     + \gamma \ln{\textbf{X}_t} + \omega \ln{\textbf{Y}_t} + \eta_{t,m} + \eta_{t,a} + \alpha_i \times \eta_{t,m} + \epsilon_i &
\end{flalign}

\noindent Exclude external control variables

\begin{flalign}
    &\textbf{M5:} \hspace{1cm}
    \ln{y_{i,t}} = \alpha_i + \beta_1 \ln{G_t} + \beta_2 \ln{S_t} + \beta_3 \ln{W_t} + \beta_4 \ln{\overline{W}_t}
     + \omega \ln{\textbf{Y}_t} + \eta_{t,m} + \eta_{t,a} + \alpha_i \times \eta_{t,m} + \epsilon_i &
\end{flalign}

\noindent Alternate treatment of demand

\begin{flalign}
    &\textbf{M6:} \hspace{1cm}
    \ln{y_{i,t}} = \alpha_i + \beta_1 \ln{D_t'} + \beta_2 \ln{S_t} + \beta_3 \ln{W_t} + \beta_4 \ln{\overline{W}_t}
     + \gamma \ln{\textbf{X}_t} + \omega \ln{\textbf{Y}_t} + \eta_{t,m} + \eta_{t,a} + \alpha_i \times \eta_{t,m} + \epsilon_i &
\end{flalign}

where $D_t'$ is the residual demand given by $D_t' = D_t - H_t + I_t$. 

\noindent Exclude wind ramp

\begin{flalign}
    &\textbf{M7:} \hspace{1cm}
    \ln{y_{i,t}} = \alpha_i + \beta_1 \ln{G_t} + \beta_2 \ln{S_t} + \beta_3 \ln{W_t} +  \gamma \ln{\textbf{X}_t} + \omega \ln{\textbf{Y}_t} + \eta_{t,m} + \eta_{t,a} + \alpha_i \times \eta_{t,m} + \epsilon_i &
\end{flalign}

\noindent Exclude time fixed effects

\begin{flalign}
    &\textbf{M8:} \hspace{1cm}
    \ln{y_{i,t}} = \alpha_i + \beta_1 \ln{G_t} + \beta_2 \ln{S_t} + \beta_3 \ln{W_t} + \beta_4 \ln{\overline{W}_t}
     + \gamma \ln{\textbf{X}_t} + \omega \ln{\textbf{Y}_t} + \alpha_i \times \eta_{t,m} + \epsilon_i &
\end{flalign}

\noindent Exclude interaction terms

\begin{flalign}
    &\textbf{M9:} \hspace{1cm}
    \ln{y_{i,t}} = \alpha_i + \beta_1 \ln{G_t} + \beta_2 \ln{S_t} + \beta_3 \ln{W_t} + \beta_4 \ln{\overline{W}_t}
     + \gamma \ln{\textbf{X}_t} + \omega \ln{\textbf{Y}_t} + \eta_{t,m} + \eta_{t,a} + \epsilon_i &
\end{flalign}

\begin{table*}[ht!]
\centering
\caption{Coefficients for the panel regression formulation for natural gas plants in CAISO for various model specifications}
\label{table:coefficients-CAISO}
\begin{tabularx}{\textwidth}{lXXXXXXXXX}
    \toprule
    \rowcolor{Gray} Variable & M1 & M2 & M3 & M4 & M5 & M6 & M7 & M8 & M9 \\
    \midrule
    $\ln{G_t}$ & 2.80*** & - & 2.75*** & 2.93*** & 2.89*** & - & 2.87*** & 2.91*** & 2.75*** \\
     & (0.07) & - & (0.07) & (0.07) & (0.06) & - & (0.07) & (0.06) & (0.07) \\
    $\ln{S_t}$ & -0.22*** & -0.10*** & - & -0.19*** & -0.24*** & -0.28*** & -0.23*** & -0.19*** & -0.21*** \\
     & (0.03) & (0.03) & - & (0.03) & (0.02) & (0.03) & (0.03) & (0.03) & (0.03) \\
    $\ln{W_t}$ & -0.20*** & -0.26*** & -0.20*** & - & -0.21*** & -0.23*** & -0.13*** & -0.22*** & -0.19*** \\
     & (0.01) & (0.01) & (0.01) & - & (0.01) & (0.01) & (0.01) & (0.01) & (0.01) \\
    $\ln{S_{\text{ext},t}}$ & 0.00 & -0.15*** & -0.16*** & 0.03 & - & 0.01 & 0.01 & -0.03 & -0.01 \\
     & (0.03) & (0.03) & (0.02) & (0.03) & - & (0.03) & (0.03) & (0.03) & (0.03) \\
    $\ln{W_{\text{ext},t}}$ & -0.02*** & -0.04*** & -0.02*** & -0.04*** & - & -0.00 & -0.02*** & -0.02*** & -0.02*** \\
     & (0.00) & (0.00) & (0.00) & (0.00) & - & (0.00) & (0.00) & (0.00) & (0.01) \\
    $\ln{D_{\text{ext},t}}$ & 0.07** & 0.61*** & 0.11*** & 0.03 & - & 0.03 & 0.06* & 0.08** & 0.10*** \\
     & (0.03) & (0.03) & (0.03) & (0.03) & - & (0.03) & (0.03) & (0.03) & (0.03) \\
    $\ln{W_{\text{ramp},t}}$ & 0.12*** & 0.20*** & 0.12*** & -0.06*** & 0.12*** & 0.11*** & - & 0.13*** & 0.11*** \\
     & (0.01) & (0.01) & (0.01) & (0.01) & (0.01) & (0.01) & - & (0.01) & (0.02) \\
    $\ln{D_t'}$ & - & - & - & - & - & 1.95*** & - & - & - \\
     & - & - & - & - & - & (0.04) & - & - & - \\
    \midrule
    R-squared & 0.84 & 0.83 & 0.84 & 0.84 & 0.84 & 0.84 & 0.84 & 0.84 & 0.82 \\

    \midrule

    Time FE & Y & Y & Y & Y & Y & Y & Y & N & Y \\
    Interaction & Y & Y & Y & Y & Y & Y & Y & Y & N \\
    \bottomrule

\end{tabularx}
\end{table*}

\begin{table*}[ht!]
\centering
\caption{Coefficients for the panel regression formulation for natural gas and coal plants in ERCOT for various model specifications}
\label{table:coefficients-ERCOT}
\begin{tabularx}{\textwidth}{lXXXXXXXXX}
    \toprule
    \rowcolor{Gray} Variable & M1 & M2 & M3 & M4 & M5 & M6 & M7 & M8 & M9 \\
    \midrule
    $\ln{G_t}$ & 1.82*** & - & 1.81*** & 1.36*** & 1.86*** & - & 1.76*** & 1.71*** & 1.69*** \\
     & (0.04) & - & (0.04) & (0.04) & (0.03) & - & (0.04) & (0.04) & (0.04) \\
    $\ln{S_t}$ & -0.03*** & 0.00 & - & 0.06*** & -0.03*** & -0.03*** & 0.02** & -0.07*** & -0.03*** \\
     & (0.01) & (0.01) & - & (0.01) & (0.01) & (0.01) & (0.01) & (0.00) & (0.01) \\
    $\ln{W_t}$ & -0.32*** & -0.24*** & -0.31*** & - & -0.34*** & -0.32*** & -0.29*** & -0.32*** & -0.30*** \\
     & (0.01) & (0.01) & (0.01) & - & (0.01) & (0.01) & (0.01) & (0.01) & (0.01) \\
    $\ln{S_{\text{ext},t}}$ & 0.01* & -0.00 & -0.00 & 0.02*** & - & 0.01* & 0.00 & 0.04*** & 0.01* \\
     & (0.01) & (0.01) & (0.01) & (0.01) & - & (0.01) & (0.01) & (0.01) & (0.01) \\
    $\ln{W_{\text{ext},t}}$ & -0.03*** & -0.10*** & -0.03*** & -0.24*** & - & -0.03*** & -0.04*** & -0.03*** & -0.03*** \\
     & (0.01) & (0.01) & (0.01) & (0.01) & - & (0.01) & (0.01) & (0.01) & (0.01) \\
    $\ln{D_{\text{ext},t}}$ & 0.05 & 1.33*** & 0.06 & 0.29*** & - & 0.05 & 0.06 & -0.01 & 0.07 \\
     & (0.05) & (0.03) & (0.04) & (0.05) & - & (0.05) & (0.05) & (0.04) & (0.05) \\
    $\ln{W_{\text{ramp},t}}$ & 0.12*** & 0.07*** & 0.10*** & -0.02*** & 0.12*** & 0.12*** & - & 0.13*** & 0.11*** \\
     & (0.01) & (0.01) & (0.01) & (0.01) & (0.01) & (0.01) & - & (0.01) & (0.01) \\
    $\ln{D_t'}$ & - & - & - & - & - & 1.81*** & - & - & - \\
     & - & - & - & - & - & (0.04) & - & - & - \\
    \midrule
    R-squared & 0.84 & 0.83 & 0.84 & 0.84 & 0.84 & 0.84 & 0.84 & 0.84 & 0.82 \\
    \midrule
    Time FE & Y & Y & Y & Y & Y & Y & Y & N & Y \\
    Interaction & Y & Y & Y & Y & Y & Y & Y & Y & N \\
    \bottomrule
\end{tabularx}
\end{table*}

\newpage

\subsection{Power plant emissions data comparison}
\label{emissions_scenarios}

\begin{figure}
    \centering
    \begin{subfigure}[b]{0.8\linewidth}
        \centering
        \includegraphics[width=\linewidth]{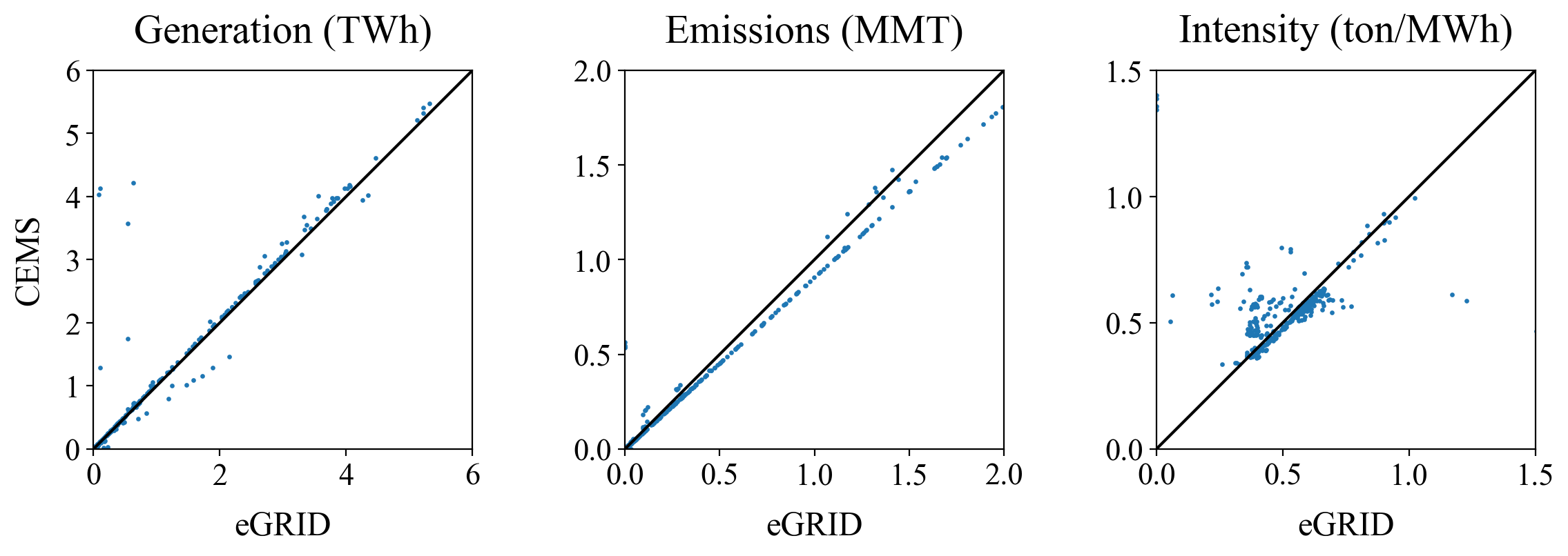}
        \caption{CAISO, Natural Gas}
        \label{subfig:CAISO_NG}
    \end{subfigure}
    
    \begin{subfigure}[b]{0.8\linewidth}
        \centering
        \includegraphics[width=\linewidth]{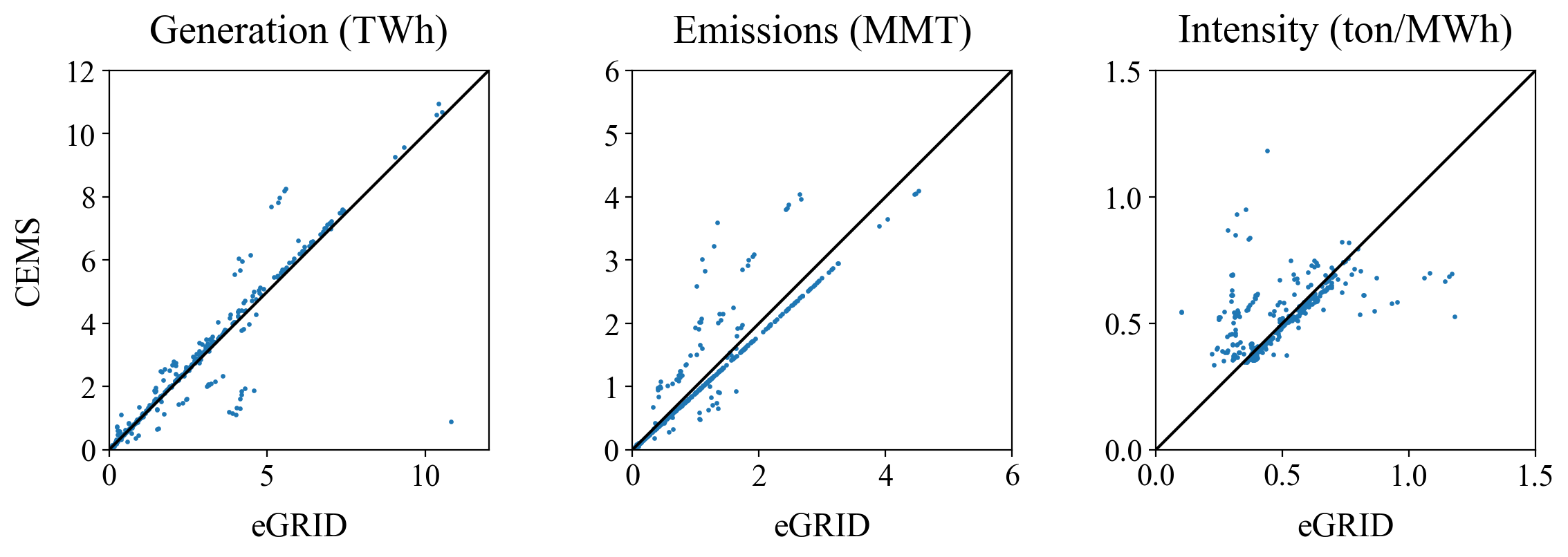}
        \caption{ERCOT, Natural Gas}
        \label{subfig:ERCOT_NG}
    \end{subfigure}
    
    \begin{subfigure}[b]{0.8\linewidth}
        \centering
        \includegraphics[width=\linewidth]{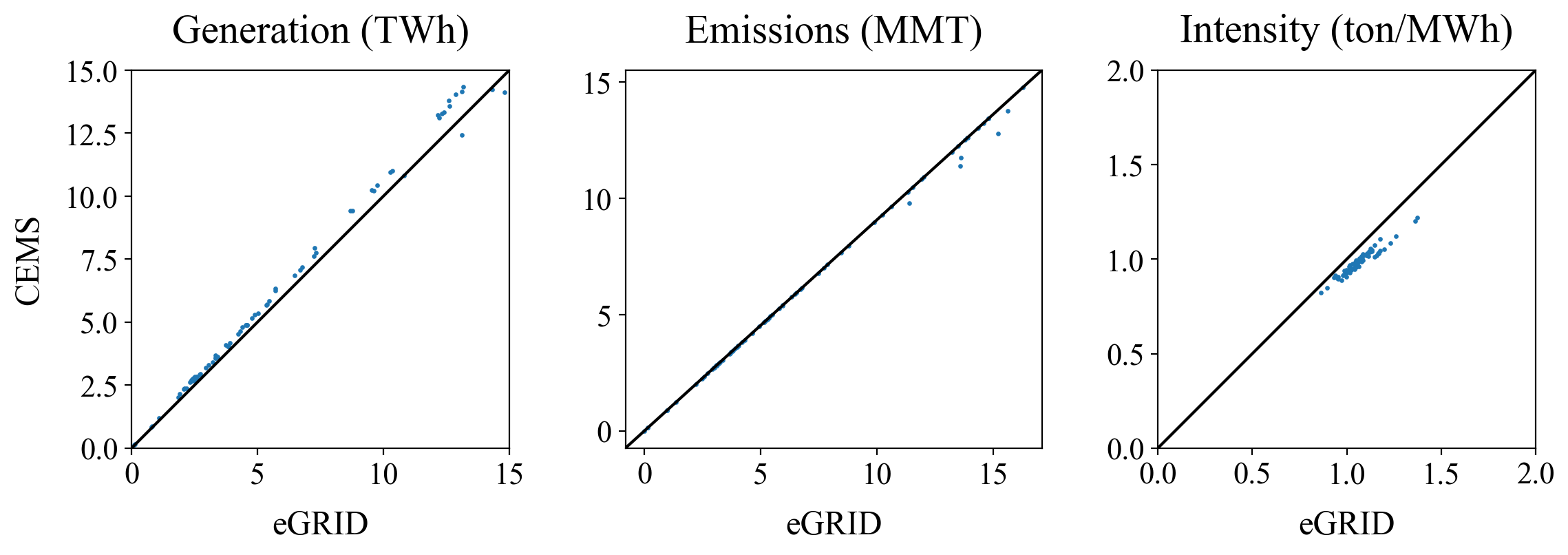}
        \caption{ERCOT, Coal}

        \label{subfig:ERCOT_COAL}
    \end{subfigure}

    \caption{Comparison of generation, emissions, and emissions intensity across CEMS and eGRID. Each point represents annual aggregated data for individual plants. If both data sources were 1:1, all points would lie on the solid black line, which is not the case. This helps us explain deviations in the scenarios where we use eGRID plant-level and BA-wide emissions intensities to compute cumulative annual emissions.}
    
    \label{fig:scnarios-comp}
\end{figure}

\end{document}